\documentclass[twocolumn]{aastex61}

\usepackage{amsmath, amsthm, amssymb}
\usepackage{bm}
\usepackage{multirow}
\usepackage{xcolor}                                                                      
\usepackage{colortbl} 
\revised{\today}
\submitjournal{ApJ}

\shorttitle{Dynamo in self-gravitating disks}
\shortauthors{Deng, Mayer \& Latter}

\begin{document}

\title{Global simulations of self-gravitating magnetized protoplanetary disks}

\correspondingauthor{Hongping Deng}
\email{hpdeng@physik.uzh.ch}

\author{Hongping Deng}

\affiliation{Center for Theoretical Astrophysics and Cosmology, Institute for Computational Science, University of Zurich, Winterthurerstrasse 190, 8057 Zurich, Switzerland}
\affiliation{Department of Applied Mathematics and Theoretical Physics, University of Cambridge, Centre for Mathematical Sciences, Wilberforce Road, Cambridge CB3 0WA, UK}

\author{Lucio Mayer}
\affiliation{Center for Theoretical Astrophysics and Cosmology, Institute for Computational Science, University of Zurich, Winterthurerstrasse 190, 8057 Zurich, Switzerland}

\author{Henrik Latter}
\affiliation{Department of Applied Mathematics and Theoretical Physics, University of Cambridge, Centre for Mathematical Sciences, Wilberforce Road, Cambridge CB3 0WA, UK}

\begin{abstract}
In  the early stages of a protoplanetary disk, when its mass is a significant fraction of its star's, turbulence generated by
gravitational instability (GI)
should feature significantly in the disk's evolution. At the same time, the disk may
be sufficiently ionised for magnetic fields to play some role in the dynamics.
Though usually neglected, the impact of magnetism on the GI may be critical, with consequences
for several processes: the efficiency of accretion, spiral structure formation, fragmentation, and the
dynamics of solids. In this paper, 
we report on global three-dimensional
magnetohydrodynamical simulations of a self-gravitating protoplanetary disk using the meshless finite mass (MFM)
Lagrangian  technique.
We confirm that GI spiral waves trigger a dynamo
that amplifies
an initial magnetic field to nearly thermal
amplitudes (plasma $\beta <10$), an order of magnitude greater
than that generated by the magneto-rotational instability alone.
We also determine the dynamo's nonlinear back reaction on the gravitoturbulent flow:
the saturated state is substantially hotter, with an associated larger Toomre parameter and weaker, more `flocculent' spirals. 
But perhaps of greater import is the dynamo's boosting of accretion via a significant Maxwell stress; mass accretion is enhanced
by factors of several relative  to either pure GI or pure MRI.
Our simulations use ideal MHD, an admittedly poor approximation in protoplanetary disks, and  
thus future studies should explore the full gamut of non-ideal MHD. In preparation for that, we exhibit
a small number of Ohmic runs that reveal that the dynamo, if anything, is stronger in a non-ideal environment.
This work confirms that magnetic fields are a potentially critical ingredient in gravitoturbulent 
young disks, possibly controlling their evolution, especially
via their enhancement of (potentially episodic) accretion.
\end{abstract}




\section{Introduction} 
\label{sec:intro}

One of the fundamental problems in planet and star formation concerns the nature of angular momentum
transport in protoplanetary disks. Not only does this process govern
the rate and nature of mass accretion on to the (proto-)star, it determines
how material is redistributed through the disk and, consequently,
the conditions for planet formation. Unfortunately, observational estimates of mass 
accretion rates in young disks rates are rather sporadic, 
but there is mounting evidence that the majority of
mass transfer, and indeed planet formation, occurs early (0.1-1 Myr) \citep[see, e.g.][]{Helled2014}. 
For example, the well established `disk luminosity problem' 
is one indication of strong early accretion \citep{Hartmann1996},
as is the fact that Type II
and older disks possess masses that are too small in comparison to those of observed
exoplanetary systems \citep{Najita2014,Manara2018}.

It is early during a protostellar disk's life (when its mass is a significant fraction of its host star)
that it is most susceptible to 
gravitational instability (GI). Estimates of disk masses suggest that 50\% of
Class 0 and 25\% of Class I disks are GI unstable \citep{Kratter2016},
while recent images of spiral structure in
some young sources (e.g. Elias 2-27, WaOph 6) are consistent with the activity of GI \citep{Perez2016,Meru2017,Huang2018}. 

On the other hand, hydrodynamical simulations of collapsing molecular cloud cores
recurrently produce massive and self-gravitating disks \citep{Vorobyov2010,Vorobyov2015,Hayfield2011}, and this holds even 
when magnetic fields are included, provided that non-ideal MHD effects are taken into account \citep{Tomida2017,Lam2019}.
In fact, most models of protostellar disk evolution posit that it is turbulence
instigated by GI that drives mass accretion during their early years \citep[e.g.][]{Durisen2007},
precisely the period in which we have evidence for the most active accretion, and possibly
planet formation. 
This provides strong motivation to fully establish theoretically the dynamics of GI.

The onset and saturation of GI in protostellar disks have been thoroughly studied
with hydrodynamic models. Magnetic fields have almost always been neglected. While it is true that protostellar disks exhibit
notoriously low ionisation fractions, there is strong numerical evidence that magnetic fields
remain dynamically important nonetheless, both in the earlier core collapse and in the later T-Tauri phases \citep[e.g.,][]{Masson2016,Turner2014}.
Simulations of the latter indicate that non-ideal MHD effects limit the magnetorotational instability (MRI)
to certain radii, but still permit significant angular momentum transport via the
formation of laminar magnetic outflows \citep[see, e.g.,][]{Bai2013, Lesur2014, Bai2014, Simon2015, Gressel2015, Bethune2017}.
Observationally,
there is some (contested) evidence of disordered fields lying primarily in the disk plane
from dust polarimetry, most notably in the cases of 
HL Tau and the class 0 object, I16293B \citep{Stephens2014,Rao2013}. Future observations of Zeeman splitting 
of CN lines by ALMA may provide 
further information about these in situ fields \citep{Brauer2017, Vlemmings2019}.
Given the prominence of non-ideal MHD in T-Tauri disks, it is natural to ask how magnetic fields alter, and
become altered by, the turbulence generated by GI, even if at the present time the ionisation profile of young disks is poorly 
constrained. 


The first direct MHD simulations of a self-gravitating disk were carried out by \citet{Fromang2004} and \citet{Fromang2005} who showed 
that MHD turbulence simply reduced the effectiveness of  GI transport. 
However, this pioneering work could only afford a rather low resolution, did not treat the
full 2$\pi$ in azimuth, nor could be run for many orbits. More recently,  
\citet{Riols2017,Riols2018} presented the first high-resolution, long-time simulations of GI turbulence and magnetic fields,
but in vertically stratified boxes. 
These revealed
several surprising results: (a) gravitoturbulence impedes and can ultimately overwhelm the MRI, when the cooling time is sufficiently low,
(b) gravitoturbulence functions as a dynamo itself, building up strong fields
even in highly resistive gas,  (c) the magnetic fields so generated can reach nearly thermal
strengths, and their back-reaction on the flow severely weakens the gravitoturbulent spiral waves, and
(d) the resulting accretion torques are enhanced,
dominated by magnetic stresses, and show oscillatory behavior.
It is clear from this work that the admixture of magnetic fields and GI produce qualitatively different outcomes than in pure hydrodynamics. But being local simulations, they raise several issues that need to be addressed. For instance, the GI dynamo occurs mainly on large (possibly global) scales, and gravitoturbulence itself possesses an inherently global character, at least in a thicker protostellar disk environment. These issues motivate the simulation of GI and MHD in \emph{global models of protostellar disks}, which is the task this paper sets itself.

We explore the GI dynamo in fully 3D global simulations 
of massive self-gravitating magnetized protoplanetary disks.
Most of our runs employ \emph{ideal MHD}, and not the full gamut of non-ideal effects prevailing in real disks.
They should be regarded as an unavoidable first step before embarking on an exploration of models employing more realistic, but also more complicated and poorly constrained, ionisation physics.
We employed the N-Body+MHD code
GIZMO \citep{Hopkins2015a,Hopkins2015b,Hopkins2016a} in its meshless finite mass (MFM) mode \citep[see also][]{Gaburov2011}.
It has been demonstrated that the MFM approach performs 
especially well in simulations of pure GI, exhibiting better conservation of angular
momentum than most competing methods, which makes possible, for example, the numerical
convergence of the cooling boundary in fragmentation studies\citep{Deng2017}. 
Moreover, while more diffusive than certain finite volume grid-based codes, it can adequately handle MHD, as shown by recent local simulation of the MRI \citep{Deng2019}.
Given that our aim is to uncover the global GI dynamo, which is large-scale (and, if anything, enhanced by diffusion),
the MFM particle method is a suitable tool for our task. It should be stressed from the outset that
our goal is \emph{not} to provide a comprehensive study of global magnetorotational turbulence, which is a challenging
problem even in grid codes, and certainly more so for a particle code such as MFM.  
We also emphasise that the numerical task is especially heavy: our best resolved runs employed
$\sim40$M particles per disk, and are thus some of the most 
expensive self-gravitating disk simulations attempted 
so far~\citep[see, e.g. hydrodynamical simulations,][]{Meru2012,Szulagyi2016}. The two GI-MHD simulations alone employed nearly 3 million core hours on the 
CPU-only partition of the Cray XC40/XC50 supercomputer {\it Piz Daint} at the
Swiss National Supercomputing Center (CSCS).

Our results can be summarised as follows. Simulations that mixed GI and MHD generated large-scale magnetic fields of a different character and strength than those sustained by pure MRI runs with no self-gravity. 
In particular, their saturated magnetic energies were roughly
an order of magnitude higher (with plasma betas significantly lower, approaching $\sim 10$), and their poloidal fields were organised around the GI spiral waves in characteristic rolls \citep[in accordance with][]{Riols2018}. We hence conclude that the GI dynamo can manifest in global disks. As a result of magnetic pressure and enhanced 
magnetic dissipation, the disk becomes hotter and thicker, while the back reaction of the dynamo via the Lorenz force degrades the spiral structure, rendering it more `flocculent'. Finally, the highly magnetised dynamo state produces a large Maxwell stress. As a consequence, the mass accretion rate in GI-MHD runs can be several times the mass accretion rate in purely hydro GI runs: magnetic fields significantly `speed up' the evolution of the disk. Finally, as a prelude to future work involving non-ideal MHD, we ran a small number of Ohmic simulations, and found that the GI dynamo is mostly unchanged in its key features. In fact it is slightly more vigorous.



 The structure of the paper is as follows: in section~\ref{sec:disk}, we describe the governing equations, numerical method, simulation setup and diagnostics. In section~\ref{sec:results} we present our results, first analysing the growth and properties of the magnetic dynamo, and second assessing its back reaction on the gravitoturbulent state. We explore the effects of magnetic diffusivity on the dynamo in section~\ref{sec:ohmic}. The caveats and outlook are discussed in section~\ref{sec:disc} and we draw our conclusions in section~\ref{sec:con}.

\section{Physical and numerical model}
\label{sec:disk}
\subsection{The governing equations}
The equations we solve are those of compressible self-gravitating MHD:
\begin{align}
  \frac{\partial \rho}{\partial t}+&\bm{\nabla}\cdot(\rho \bm{v})=0,\\
  \frac{\partial \bm{v}}{\partial t}+\bm{v}\cdot\bm{\nabla}\bm{v}&=-\frac{1}{\rho}\bm{\nabla}(P+\frac{B^{2}}{8\pi})+\frac{(\bm{B}\cdot\bm{\nabla})\bm{B}}{4\pi\rho} -\bm{\nabla}\Phi, \label{eq:eom} \\
  \frac{\partial \bm{B}}{\partial t}&=\bm{\nabla}\times(\bm{v}\times\bm{B})+\eta \bm{\nabla}^{2}\bm{B},  \\
  \frac{\partial U}{\partial t}& +\bm{\nabla}\cdot (U \bm{v})=-P\bm{\nabla}\cdot\bm{v}-\frac{U}{\tau_c},
\end{align}
where  $\rho$, $U$, $P$, and $\bm{v}$ represent the density, internal energy, gas pressure, and velocity respectively; $\bm{B}$ is the magnetic field and $\eta$ is the magnetic resistivity. We focus on ideal MHD, where $\eta=0$, in the paper though present exploratory simulations with magnetic resistivity in section \ref{sec:ohmic}. The ratio between the gas pressure and magnetic energy, $\beta\equiv P/(B^{2}/8\pi)$, is a widely used dimensionless measure of the magnetic field strength. $\Phi$ is the sum of the gravitational potential of the central object and the gravitational potential induced by the disk itself, $\Phi_s$, which satisfies the Poisson equation
\begin{equation}
\nabla^{2}\Phi_{s}=4\pi G \rho.
\end{equation}

We assume an ideal gas equation of state (EOS),
\begin{equation}
P=(\gamma -1)U,
\end{equation}
with $\gamma=5/3$. We adopt an \emph{ad hoc} cooling time scale that equals the local orbital period of
fluid elements, ie, $\tau_c =2\pi/\Omega(r)$ \citep{Gammie2001}. With this cooling rate the disk will not fragment \citep{Deng2017}, 
but the induced spiral pattern should be strong enough to drive a dynamo according to 
local simulations \citep{Riols2017,Riols2018}. We also note that in the MHD simulations without self-gravity, designed to study the MRI, we do not
employ any cooling. Material out-flowing from the simulation will cool the MRI disk once the disk inflates.

\subsection{Numerical method and basic set-up}
\label{sec:code}
We use the N-Body + MHD code, GIZMO\citep{Hopkins2015a,Hopkins2015b,Hopkins2016a,Hopkins2017}, in meshless finite mass (MFM) mode \citep{Gaburov2011}. The GIZMO code solves for the
disk self-gravity by employing a tree algorithm drawn from GADGET3 \citep{Springel2005}. We used the conservative and adaptive gravitational softening of \citet{Price2007},
and employed the Wendland C4 kernel with 200 neighbours \citep{Dehnen2012}.
The divergence of magnetic field is kept to low levels by the aggressive constrained gradient flux cleaning algorithm \citep{Hopkins2016a}, but see below.
The MHD module has
been tested in~\citet{Hopkins2015b} and, in addition, \citet{Deng2019} showed that GIZMO MFM describe the local MRI adequately for some 50 orbits,
provided sufficient resolution was deployed. It is noteworthy that comparable SPH MHD schemes struggled with the MRI, and in fact typically grew unphysically
strong toroidal fields. 

We performed three types of simulations, which are summarised in table ~\ref{t:simulations}: (1) global MHD simulations without self-gravity (run labels have prefix `MRI'); (2) global self-gravitating
simulations without magnetic fields (prefixed with `grvhd'); (3) global MHD simulations with self-gravity (`grvmhd'). 
Our main focus here, of course, is the third class of simulations, which is the least well explored, but the other two are necessary as they provide points of comparison.

We simulate disks exhibiting a radial range of $5<R<25$ AU orbiting a solar mass star. Hence the outer rotation period (ORP) at 25 AU is 125 yrs, and is sometimes used
as a time unit. More generally, however, we take 1 solar mass, $1$ AU, $1/2\pi$ yr and 1 Gauss as the mass, length, time and magnetic field strength units respectively. 
 The central star is modeled as an active sink particle with a sink radius of 5 AU. Gas particles reaching the sink radius are deleted and their mass  and momentum are
 added to the star to ensure mass and momentum conservation. We apply outflow boundary condition by clipping any particle whose smoothing length is larger than 5AU. 
 In grvmhd1,  this yields a density floor about $8\times 10^{-16} $g/cm$^{3}$  which is 4 orders of magnitude smaller than the mid-plane density at 25 AU.

It must be conceded from the outset that the weak field limit, in particular, is polluted (as in all particle codes) by small-scale (resolution dependent) noise,
arising from insufficient div(B) cleaning \citep{Deng2019}. Throughout our simulations the domain averaged dimensionless divergence $\langle |h*\nabla\cdot B/B| \rangle $ is kept to $\sim 10^{-3}$,
where the angle brackets indicates a domain average and $h$ is the resolution length \citep{Deng2019}. But despite this relatively low value, the persistent
deviation from solenoidality introduces an artificial magnetic diffusion and low-level magnetic activity on the smallest scales. This additional numerical diffusivity
may explain some of the decaying MRI behaviour witnessed in \citet{Deng2019}; it also makes challenging the simulation of weak fields, and the estimation of (kinematic) 
dynamo growth rates. We hence limit ourselves to stronger field initialisations, and keep in mind the enhanced numerical resistivity exhibited by these simulations. 
It is worth stressing here that this additional diffusion, and its deleterious effects on the
MRI, are of secondary concern to us: our goal in this paper is to describe the GI dynamo.
Our MRI simulations only serve as a point of comparison, and to demonstrate that the GI
dynamo is \emph{not} the MRI.

\begin{deluxetable}{c|c|c|c|c|c}[ht!]
  \tablenum{1}
  \tablecaption{Disk simulations \label{t:simulations}}
 \tabletypesize{\scriptsize}
  \tablewidth{\textwidth}
  \tablehead{
    \colhead{Run label} & \colhead{Physics} & \colhead{Disk mass} & \colhead{Particles} & \colhead{$\tau_c\Omega$} &\colhead{Run time}}
  \startdata
  MRI-lr      &   MHD      & $0.07M_{\odot}$ & 22M & $\infty$ & 4.5 ORPs  \\
  MRI-hr      &   MHD     & $0.07M_{\odot}$ & 44M & $\infty$         & 2 ORPs \\
  grvhd1      &   SG      & $0.07M_{\odot}$ & 2M  &    $2\pi$         &        10 ORPs \\
  grvhd2       &   SG     & $ 0.13M_{\odot}$ & 2M &   $2\pi$          &      10 ORPs \\
  grvmhd1    &    SG+MHD       & $ 0.07M_{\odot}$ & 35M   &     $2\pi$           & 7 ORPs \\
  grvmhd2      &  SG+MHD       & $ 0.13M_{\odot}$ & 35M   &  $2\pi$        & 6 ORPs\\
  \enddata
  \caption{List of main production runs and their attributes. The acronym SG stands for `self-gravitating'. ORP refers to
  `outer radius orbit'.
  The number of particles and disk mass in the table take the values at the point we apply our diagnostics. 
  The relaxation  stage of grvmhd1 is not included (see section~\ref{sec:grvmhdic} and figure \ref{fig:me}).
  MRI-hr starts from the saturated state of MRI-lr (see section \ref{sec:mriic})
   and grvmhd2 starts from the saturated state of grvmhd1 to save computational resources.}
\end{deluxetable}

\subsection{Initialisation of disk models }
\label{sec:mod}
We employed global disk models similar to those in \citet{Lodato2004}.
But due to the different nature of the the three types of simulations (see table \ref{t:simulations}) the initial conditions are prepared differently.
We start from a  $\approx 0.1 M_{\odot}$ disk, with 
initial surface mass density and temperature profiles obeying $\Sigma \propto R^{-1}$ \citep{Bate2018} and $T \propto R^{-1/2}$ (vertically isothermal), respectively.
The initial temperature is normalized so that at the outer edge of the disk the Toomre $Q$ parameter equals 2. 
We generate the particle distribution through Monte Carlo sampling and then relax it to the hydrodynamical equilibrium state described above. 

In the following subsections we detail the sequence of moves to generate the required initial condition in each case. But as a general rule, we obtain
higher resolution simulations from lower ones via a particle splitting technique, which is
mass and momentum conserving;
see Appendix E in \citep{Hopkins2017}.
We do not apply particle splitting on the fly to avoid numerical instabilities. Instead, we stop the simulation and
restart it from the re-sampled initial condition. To obtain disks of larger mass, the particle splitting can be made non mass-conserving.
And due to loss of particles from the domain, the mass of the disk can be reduced from the $0.1 M_{\odot}$ initially put in.

\subsubsection{The pure MRI runs}
\label{sec:mriic}

In the base simulation, dubbed ``MRI-lr", we
used 25 million particles to sample the initial disk model (section \ref{sec:mod}) and initialised the vertical field, $B_z=0.001sin(2\pi \phi)$ after the particles are relaxed to a glassy configuration \citep[see also][]{Deng2019}.  The weak seed field grows exponentially. 
We note that in the linear growth stage a small 
fluctuation is observed, possibly caused by numerical noise or by a transient growth phase (see figure \ref{fig:me}). 
When the MRI turbulence is fully developed within 20AU, namely within the main disk body, we split the particles and rerun the simulation further. The high resolution simulation is named MRI-hr. 

\subsubsection{The pure GI runs}
Since the resolution requirements for hydrodynamical self-gravitating disks are not so stringent, compared to disks with  MHD turbulence, 
2 million particles are already enough to correctly model mass and angular momentum transport via gravito-turbulence~\citep[see][]{Cossins2009} 
and fragmentation~\citep{Deng2017}. We start directly from non-self-gravitating equilibrium disk (section \ref{sec:mod}) and let it relax to a marginally unstable, turbulent state with $\tau_c=2\pi/\Omega$.

\subsubsection{The grvmhd runs: GI+MHD}
\label{sec:grvmhdic}

 We start  from a low resolution (11M particles) non-cooling, non-self-gravitating equilibrium disk with a positive pure toroidal field, $\beta=25$, everywhere. 
 Next we add a modest cooling rate with $\tau_c=8\Omega^{-1}$ to avoid spurious fragmentation \citep{Deng2017} for 2 ORPs so that the spiral structure is fully established in the disk. 
 We then split all the gas particles by a factor of 2 once and switch to the the desired cooling rate, $\tau_c=2\pi/\Omega$. We run it for 1.6 extra ORPs. 
 By this time, the initial net-toroidal flux has been expelled from the disk now leaving a zero-net-flux disk similar to published MRI simulations with initial toroidal fields ~\citep[see, e.g.][]{Fromang2006,Flock2011}. Finally, we split the particles again (reaching $\sim$ 40M  particles) to resolve small scale turbulence and run the simulation further (see figure \ref{fig:me}).  Resolution tests in Appendix~\ref{sec:res-test} show that $\sim$ 40M particles give converged time-averaged quantities in the saturated turbulence. The simulation starting from the last re-sampling state is what we identify as the grvmhd1 simulation in table \ref{t:simulations}. The grvmhd2 model's initial condition is prepared from a grvmhd1 snapshot taken at 160 yrs by doubling the particles' mass while keeping the Toomre \emph{Q} constant (quadruple the specific internal energy). 

\subsection{Diagnostics}
\label{sec:diag}

In order to analyze the numerical results we define various averages of a quantity X.
\begin{equation}
\langle X \rangle=\frac{\int_{V}\rho X dV}{\int_V \rho dV},
\end{equation}
is the density-weighted average, where $V$ denotes the volume of the computational domain. In particle codes, it is more natural to compute this density-weighted average than the volume averaged one to avoid bias towards the under-resolved low density regions. In GIZMO this average is accomplished through
\begin{equation}
\langle X \rangle=\frac{\sum_i m_i X_i}{\sum_i m_i}=\frac{\sum_i X_i}{N}=\overline{X},\label{eq:davr}
\end{equation}
where $N$ and $m_i$ are the number of particles and the mass of the $i$th particle (particles have equal masses here). The density-weighted average $\langle X \rangle$ equals the direct arithmetic average $\overline{X}$.

We also can calculate the pure \emph{volume average}, in order to best compare with previous work. This is accomplished by adding a weighting factor $1/\rho_i$ to equation \ref{eq:davr}, i.e.,
\begin{equation}
\langle X \rangle_V=\frac{\sum_i  X_i/\rho_i}{\sum_i 1/ \rho_i},
\end{equation}
where $\rho_i$ is the mass density of the $i$th particle. 

Alongside these  are azimuthal and vertical averages:
\begin{equation}
    \langle X \rangle_{\phi}= \frac{\int_0^{2\pi}\rho X d\phi}{\int_0^{2\pi}\rho d\phi}, \qquad 
     \langle X \rangle_{z} = \frac{\int_{-\infty}^{\infty}\rho X d\phi}{\int_{-\infty}^{\infty}\rho dz},
\end{equation}
which can be combined into the double average $\langle X \rangle_{\phi z}$. Volume versions of this can also be defined.
And, finally, a temporal average 
\begin{equation}
\langle X \rangle_t(R,\phi,z) = \frac{1}{\Delta T}\int_T^{T+\Delta T} X dt,
\end{equation}
where the average takes place between times $t=T$ and $t=T+\Delta T$. Often in what follows the subscript in the average will be dropped if the context makes thigns clear.

We also computed the 2-dimensional  Toomre \emph{Q} \citep{Toomre1964},
\begin{equation}
  Q\equiv \frac{\langle c_{s} \rangle_{z} \kappa}{\pi G \Sigma},\label{eq:q2d}
  \end{equation}
  where $c_s$ is the (time and space dependent) sound speed, computed from the ideal equation of state, and the
  surface density is simply $\Sigma=\int \rho dz$.


In the paper we will quantify the transport of angular momentum via the exertion of a stress, comprising
the sum of the Reynolds stress $H_{r\phi}$, Maxwell stress $M_{r\phi}$ and the gravitational stress $G_{r\phi}$.
Here $H_{r\phi}=\rho \delta v_{r} \delta v_{\phi}$, $M_{r\phi}=-B_{r}B_{\phi}/4\pi $ and $G_{r\phi}=g_{r}g_{\phi}/4\pi G$ with $\delta v_{r}, \delta v_{\phi},g_{r},g_{\phi}$ denoting the radial/azimuthal velocity/gravitational acceleration fluctuations \citep{Lynden1972}. 
It is also common to introduce the alpha associated with local transport associated with these stresses:
\begin{equation}
\alpha=\langle H_{r\phi} + M_{r\phi} + G_{r\phi}  \rangle_V /\langle P \rangle_V.
\end{equation}
Both full domain or $\phi z$ averages can be used in the definition of $\alpha$. 

It is sometimes useful to calculate the azimuthal 
power spectrum of the density and (density weighted) magnetic energy.
A radial interval of the disk domain is divided annuli of fixed width, $\Delta R=0.5$ AU,
and in each annulus we compute the azimuthal Fourier transform of
the midplane volume density \citep[see, e.g.][]{Cossins2009}.
Within a given annulus the $m$th mode amplitude is
\begin{equation}
\Sigma_m=\frac{1}{N_\text{ann}}\left \vert \sum\limits_{k=1}^{N_\text{ann}} e^{-im\phi_k} \right \vert,
\end{equation}
where $\phi_k$ is the azimuthal angle of the $k$th particle and $N_\text{ann}$ is the number of particles in this annulus. Similarly, we can calculate the $m$th mode of the magnetic energy:
\begin{equation}
B^{2}_m=\frac{1}{N_\text{ann}}\left \vert \sum\limits_{k=1}^{N_\text{ann}}B^{2} e^{-im\phi_k} \right \vert.
\end{equation}
These quantities are subsequently time-averaged over some interval.

\begin{figure}[ht!]
  \epsscale{1.25}
\plotone{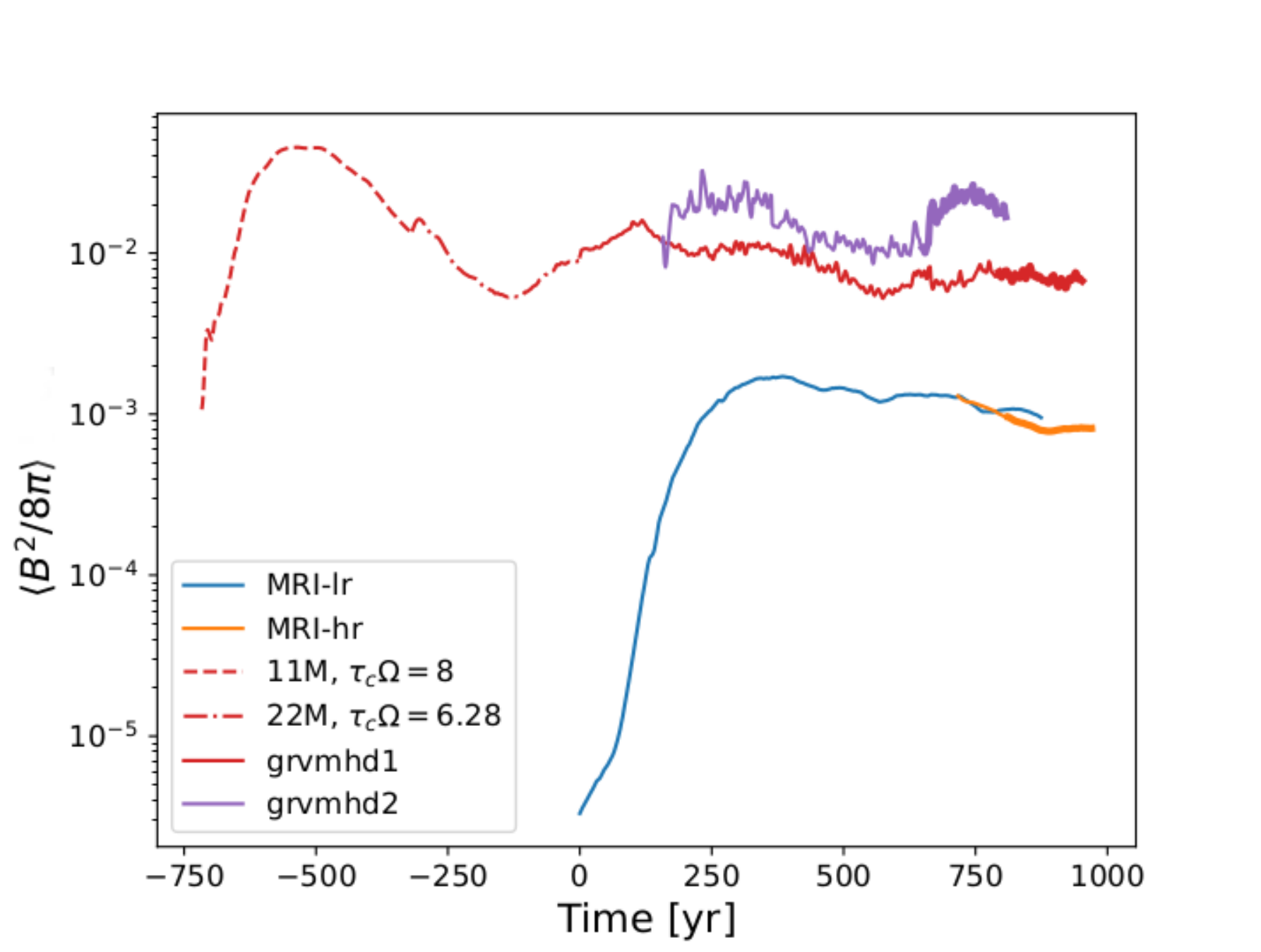}
\caption{The averaged magnetic field energy (in code units) in the disk trunk (10-20 AU) during the whole simulation. The initial relaxation for grvmhd1 is denoted by dash and dot-dash lines. Grvmhd2 starts from the saturated state of grvmhd1 (see section ~\ref{sec:grvmhdic}) and shows a transitional growth of $m=4$ mode spirals and field strength. The thick lines indicate the last 1000 code time units evolution during which we do the time-averaging. \label{fig:me}}
\end{figure}


\begin{figure*}[ht!]
\epsscale{1.2}
\plotone{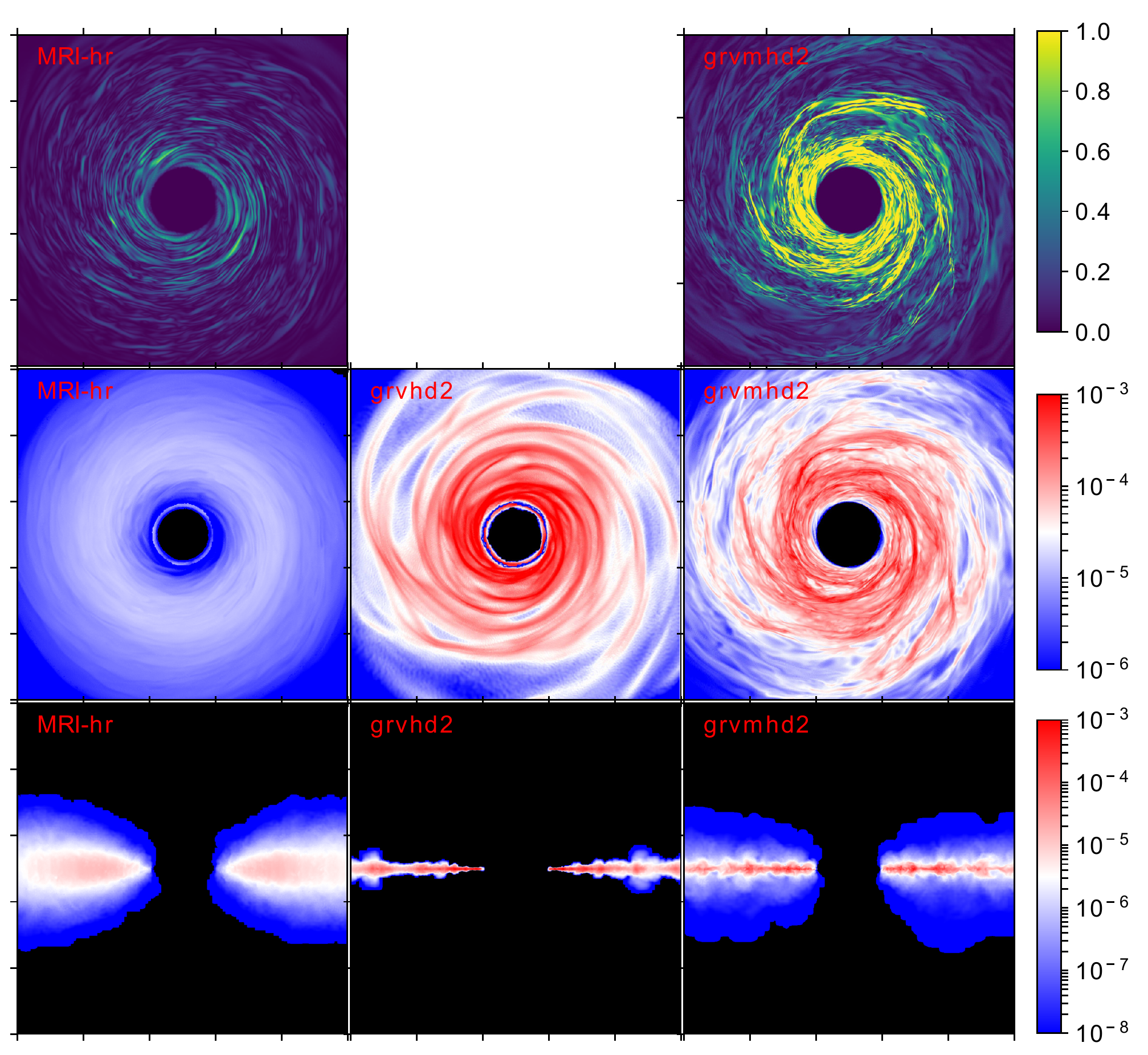}
\caption{Face-on and side-on color-coded maps of magnetic field strength and density in code units. The top row is magnetic field, the middle row is midplane density, and the bottom row is density.
The box size is 50 AU per side. The grvmhd2 run exhibits stronger magnetic fields than the MRI-hr simulation, and they are correlated with the spiral density waves. Compared to grvhd2,  grvmhd2 exhibits more flocculent spiral density waves and a much more more extended  disk atmosphere. Finally, as a test, we turned off the MHD module in grvmhd2 and grvmhd2 gradually collapse to a razor-thin state similar to grvhd2.\label{fig:compile}}
\end{figure*}

\section{Results}
\label{sec:results}
 
 Our main goal is to simulate magnetic field generation
 and saturation in GI turbulent flows. But to best understand what is going on, we also present purely hydrodynamic GI runs, so as to (a) exhibit the turbulent flows that initially give rise to this field generation, and (b) provide a point of comparison for the saturated dynamo flows, in which strong fields have reacted back on the turbulence (via the Lorenz force) and altered it. We also present pure MRI runs, with no GI. These exist to demonstrate that the dynamo fields are different in magnitude and character to those generated by the MRI. We emphasise that our aim is not
 a comprehensive study of the MRI and how well GIZMO performs in describing it; but rather to prove that the dynamo we see is \emph{not} the MRI.
 
 Rather than treating each set of simulations separately, we break up the results into (a) a descriptive section that concentrates only on the magnetic field properties of the simulations, contrasting, in particular, the MRI runs with GI-MHD runs, (b) a section that investigates the back reaction of the dynamo on the GI turbulent flow, (c) a section that undertakes some analysis of the GI dynamo, (d) a closer look at the transport of angular momentum, accretion, and outflows, and finally (e) a short section discussing some preliminary non-ideal MHD runs.


  In figure \ref{fig:me}, the time-evolution of the magnetic energy is plotted for the MHD runs. To give an impression of the different types of turbulent structures and magnetic fields, we have also plotted in  \ref{fig:compile}, the midplane density, midplane magnetic field strength, and edge-on volumetric density of the high resolution MRI simulation MRI-hr (left), the pure GI simulation grvhd2 (middle), and the GI-MHD simulation grvmhd2.


\subsection{Magnetic fields in MHD simulations}

In this subsection we focus on the nature of the magnetic
fields observed in our pure MRI and GI-MHD simulations.
Our main diagnostics will be magnetic energy, morphology and spectra, and the characteristic
dynamo patterns in the azimuthal flux witnessed in most
MRI simulations.

Note that the GI-MHD simulations are started with pure toroidal fields with $\beta=25$, in contrast to the pure MRI runs which begin with an azimuthally varying vertical field. However, grvmhd1 has already lost any memory of the initial toroidal field after our initial relaxation process (it has been expelled).  This is similar to what has been reported in global MRI 
simulations with initial toroidal fluxes \citep{Fromang2006,Beckwith2011,Flock2011}. Thus comparison of the two sets of simulations remains valid.



\subsubsection{Energetics}
\label{sec:fields}


\begin{figure}[ht!]
\epsscale{1.2}
\plotone{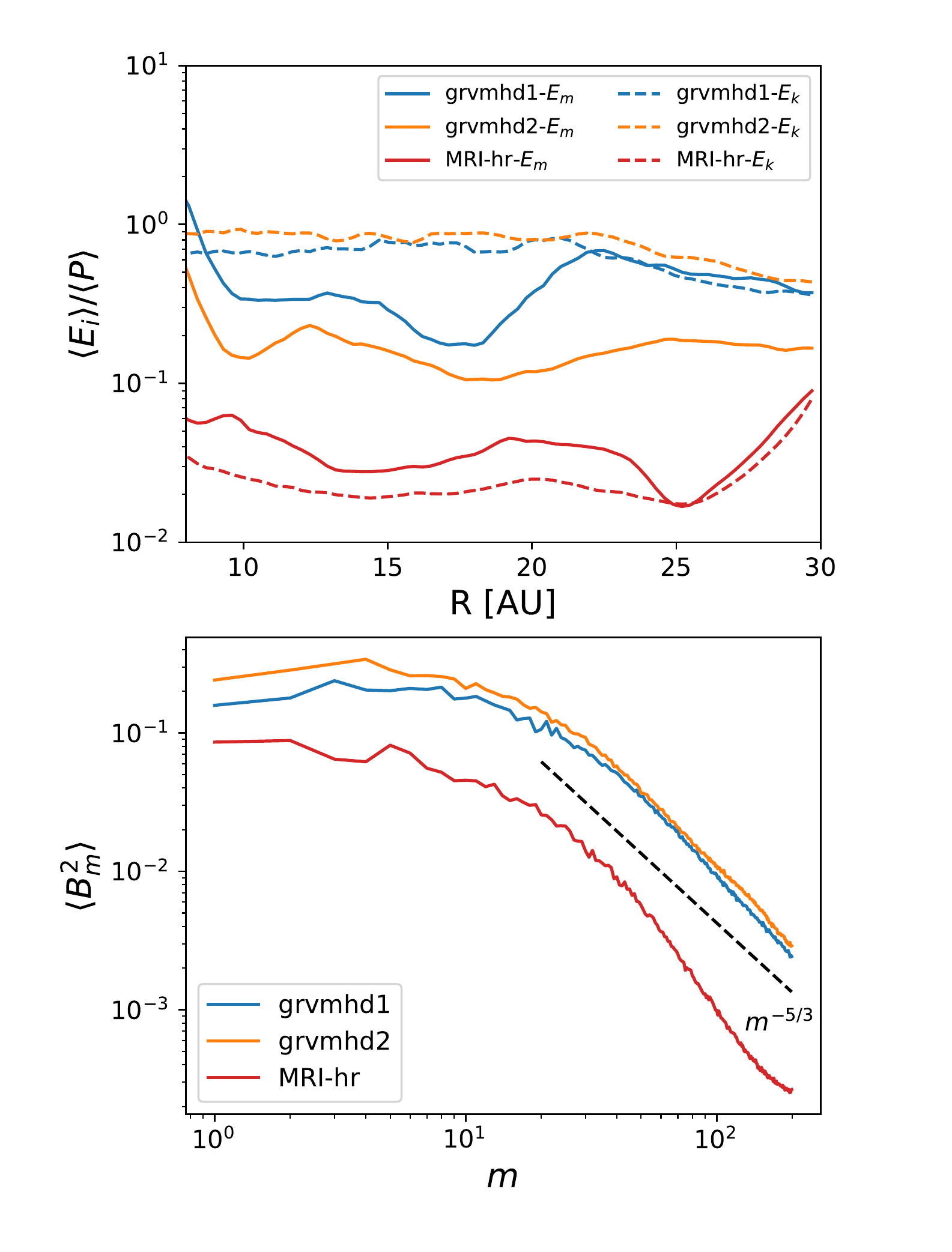}
\caption{Time averaged magnetic energy ($E_m$) and kinetic energy ($E_k$) (normalised to the averaged gas pressure) as a function of radius for the MHD simulations.\label{fig:ekem} }
\end{figure}

Perhaps the most telling difference between the pure MRI and GI-MHD runs is the magnitude of the saturated magnetic energy, as shown in figure \ref{fig:ekem}. Both grvmhd1 and grvmhd2 support much stronger magnetic energies than the MRI-hr simulation, with $\langle \beta \rangle \sim 4 $ and $\langle \beta \rangle \sim 7$ in the disk trunk (10-20 AU) respectively. The magnetic energy and kinetic energy in MRI-hr is only a few percent of the gas pressure in accord with previous global simulations~\citep{Fromang2006,Parkin2013}. This striking difference is also observed in local simulations \citep{Riols2017,Riols2018}, and is perhaps the best piece of evidence we have that magnetic energy production in GI runs is fundamentally different to the MRI.

Inspection of how the energy is partitioned between magnetic and kinetic energy in our GI-MHD simulations (figure~\ref{fig:ekem}) reveals that the kinetic energy dominates the magnetic energy, the reverse to what occurs in the MRI. 
This is perhaps as expected, given the relative strength of gravitoturbulence. The ratio of magnetic to kinetic energies
is similar to that observed in local boxes though there is some dependence on total disk mass (a quantity difficult to measure in local simulations). It is likely that this dependence is related to differences in effective numerical diffusion  \citep[see][]{Riols2018}. More notable is that, relative to the average thermal energy, both magnetic and kinetic energies are larger in the global simulations \citep[compare with Table 1 in ][]{Riols2017}.
Moreover, there is a slight dependence in the global kinetic energy on disk mass (something difficult to model
in local simulations); these discrepancies probably indicate
shortcomings of the local model.

\subsubsection{Morphology and spectra}

Moving on from averaged quantities to the morphology of the
field we can discern additional differences between the
MRI and GI-MHD simulations. In the top row of 
Figure \ref{fig:compile} the midplane magnetic field strength 
is plotted for the runs MRI-hr and grvmhd2, putting aside
the greater strength of the fields in the latter (discussed in the previous subsection), we observe that both simulations exhibit spiral structure, with perhaps the MRI run producing tighter and less coherent spirals. If we next turn to the second row, which shows the midplane density of the flows, it is clear that the dominant magnetic structures in grvmhd2 are correlated with density structures - suggesting a close connection between the two, as discussed at length in \citet{Riols2018}. In contrast, the MRI simulations support only minor perturbations in density and these do not appear dynamically significant. 

\begin{figure}[ht!]
\epsscale{1.2}
\plotone{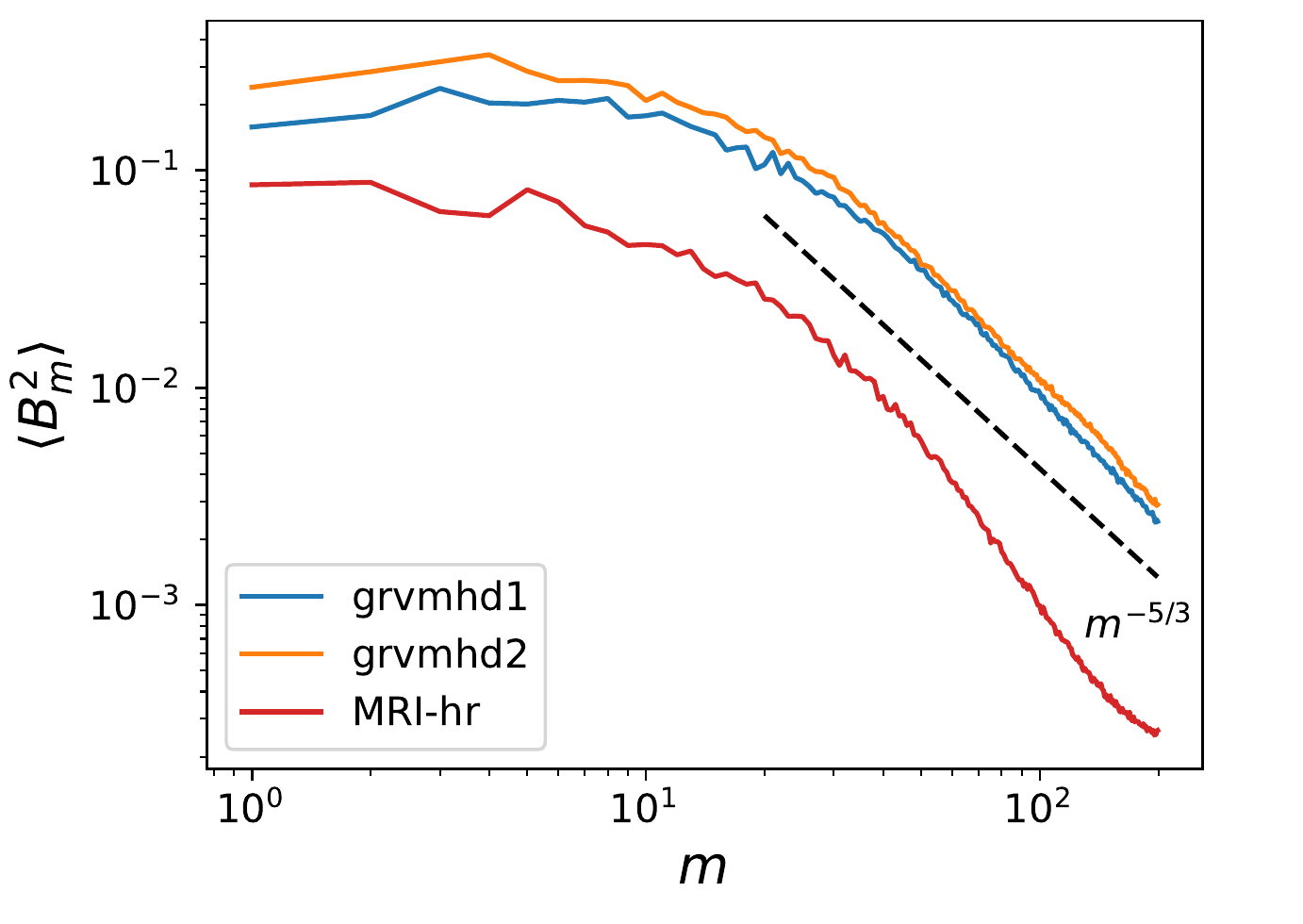}
\caption{Space (10-20AU) and time averaged (see figure~\ref{fig:me}) magnetic energy power spectrum for the MHD simulations. See figure \ref{fig:Rm100} for the effects of numerical/physical dissipation on the power spectra.  \label{fig:bm}}
\end{figure}

To obtain a more quantitative sense of the field structure, we calculate the power spectrum for the magnetic energy $B_m^2$ (see section \ref{sec:diag}).
We calculate $B_m^2$ for 10 equally spaced annuli ranging from 10 to 20 AU with width $\Delta R=0.5$ AU, noting that it varies only mildly with $R$. We then normalize the power spectrum in each bin to the local averaged magnetic energy $B^2_0$ and average the power spectrum at different radii and over time. The results are plotted in figure \ref{fig:bm} for runs MRI-hr, grvmhd1, and grvmhd2. 

The MRI magnetic energy power spectrum agrees well with the previous grid code simulations of  \citet[][Figure 12 ]{Flock2011}, 
showing a flat spectrum when $m<5$. In grvmhd1 and grvmhd2, the structure in the magnetic energy is closely related to the spiral density waves, with power peaking at $m=3$ and $m=4$ respectively, which are also the peak $m$s for the surface density spectra.

On smaller scales, the power spectrum in both GI-MHD runs shows a hint of a $m^{-5/3}$ scaling, but only on the very limited band $30<m<100$. This may, or may not, indicate the beginnings of an inertial range, with energy input at the large spiral-wave dynamo scale and then cascading to the (numerical) dissipative scales. What is less in doubt is the difference with the MRI spectrum, which shows a steeper drop in this range and no obvious scaling law, in agreement with most other local and global simulations.

\subsubsection{ Dynamo cycles}


\begin{figure}[ht!]
\epsscale{1.2}
\plotone{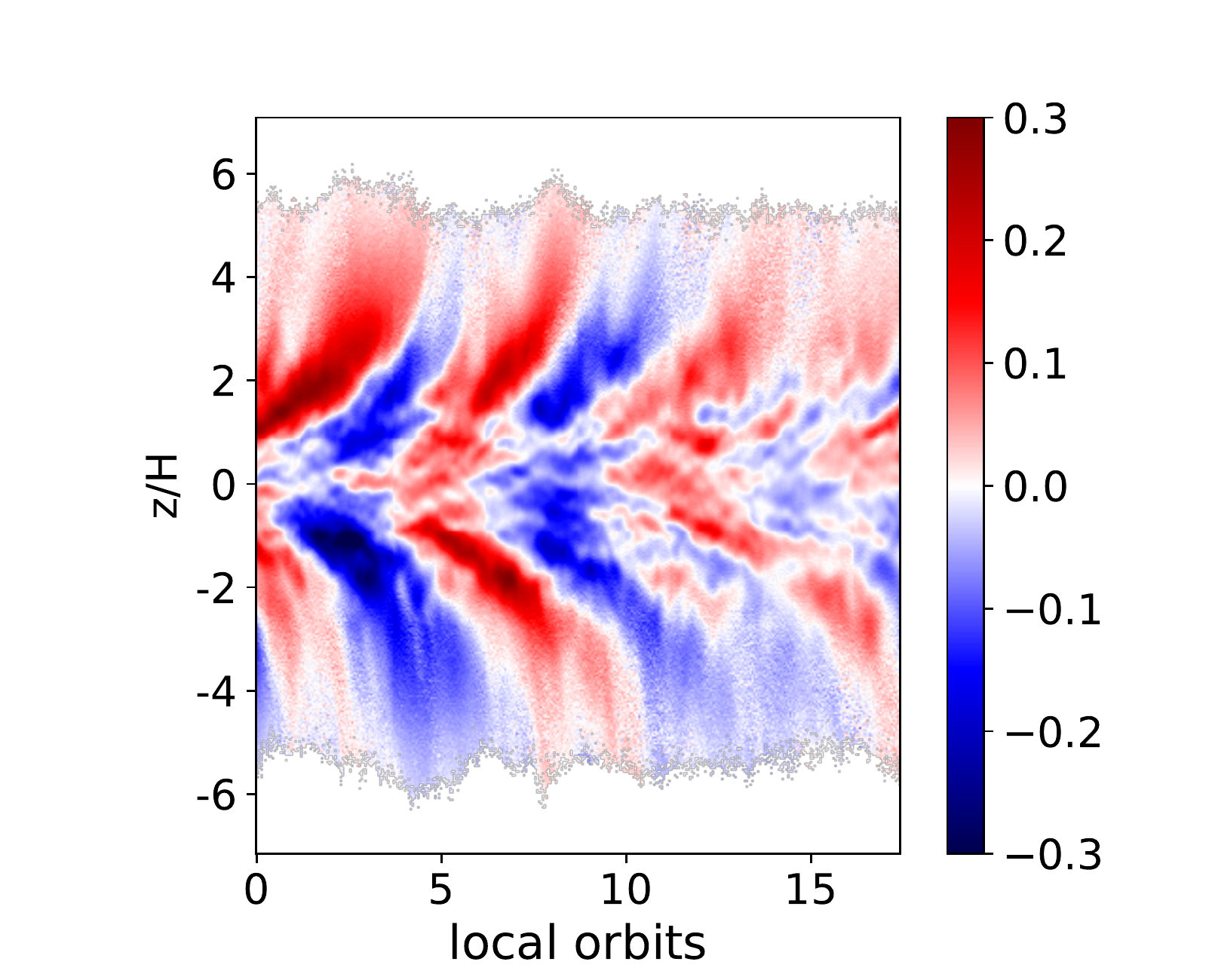}
\caption{The azimuthally averaged $\langle B_\phi \rangle$, in Gauss, between 9.5-10.5 AU in the MRI-lr simulation (250-800 yrs in figure \ref{fig:me}). MRI-lr is used here to provide long term statistics despite of the lower resolution compared to MRI-hr (see figure \ref{fig:Rm100} for numerical/physical dissipation strength).  \label{fig:fly}}
\end{figure}


 MRI turbulence shows a characteristic periodical polarity change of the azimuthally averaged toroidal fields, i.e., the so-called \emph{butterfly diagram} in both local \citep{Miller2000,Simon2015} and global simulations~\citep{Flock2011}. 
 Our MRI runs also exhibit the same temporal evolution of the toroidal field. As an example, we plot in \ref{fig:fly} $\langle B_\phi \rangle_{\phi} $ between the radii 9.5 and 10.5 AU for run MRI-lr (local orbits and disk scale height are calculated at 9.5 AU). The vertical extent of the plot is $\pm$10 AU (see also figure \ref{fig:compile}). As is clear, the polarity changes about every 8 orbits, and thus agrees with previous studies~\citep[see, e.g.][]{Flock2011}. The reproduction of this generic feature of the MRI dynamo gives us some confidence that our global simulations can describe the MRI up to some level of accuracy for some period of time \citep[see also][]{Deng2019}.



\begin{figure*}[ht!]
\plotone{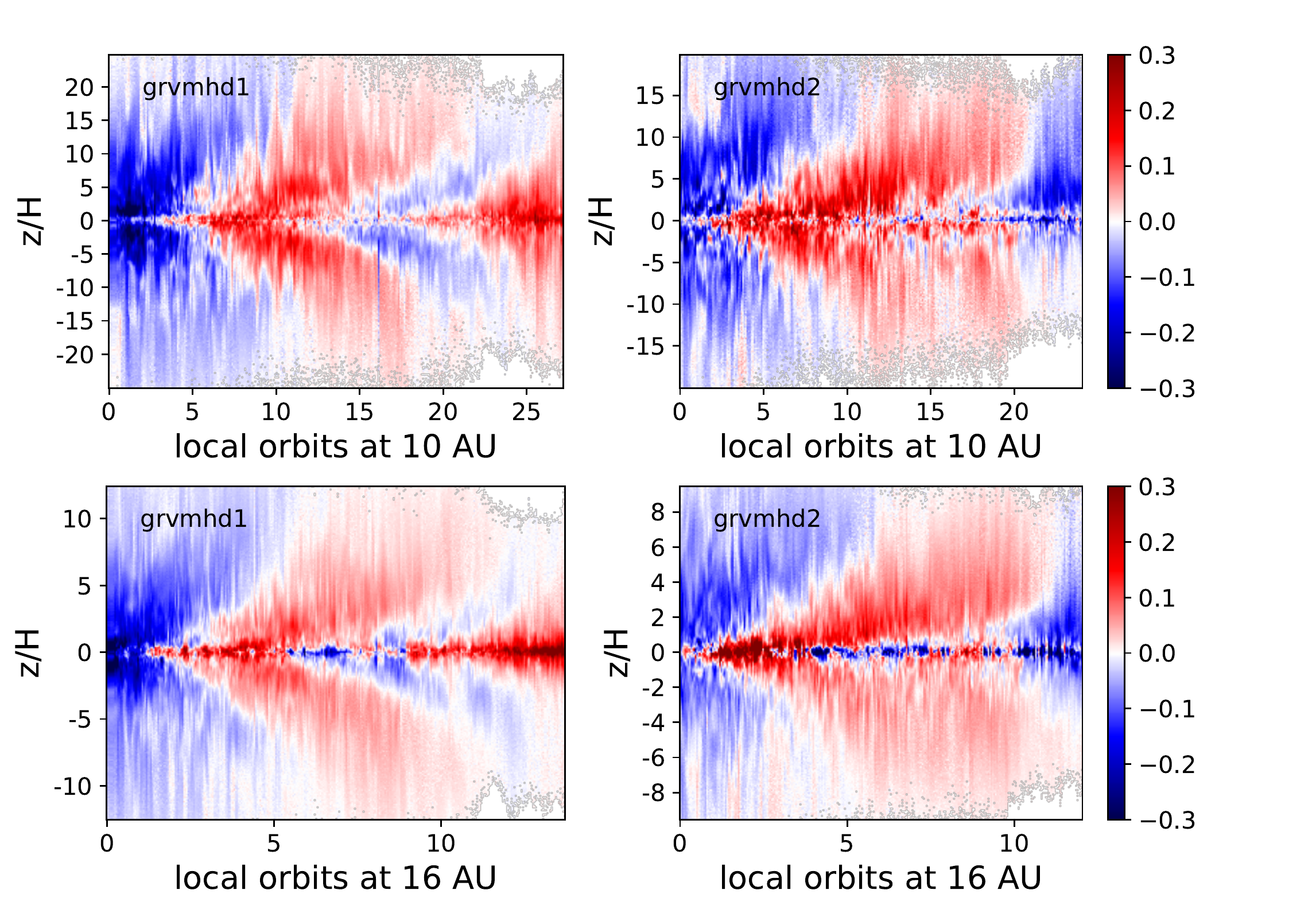}
\caption{The azimuthally averaged $\langle B_\phi \rangle$, in Gauss, between 9.5-10.5 AU (top panels)  in grvmhd1/2 and 15.5-16.5 AU (bottom panels) with shared colorbar. The local orbits and disk scale height is calculated at 10/16 AU and the vertical limits of the plots are $\pm$10 AU (see also figure \ref{fig:compile}).  The polarity of the toroidal fields changes occur on longer time scale than that in MRI (see figure \ref{fig:fly}). This time scale is independent of the local dynamic time scale. \label{fig:flux}}
\end{figure*}

 We  next show the temporal evolution of the  azimuthally averaged toroidal fields of the GI-MHD runs grvmhd1 and grvmhd2. These are plotted in figure ~\ref{fig:flux}. Both simulations show polarity changes but these occur on longer time scales; in fact, in grvmhd2 we see only one reversal, hence we cannot claim that the process is periodic. The polarity changes occur at about 470/1260 yrs in grvmhd1/2, respectively, regardless of the local dynamical time scale. It is not impossible that a significantly modified MRI is persisting on some level in these runs, but it is far more  likely that these polarity shifts are driven by the GI dynamo, and not the MRI.
 
 Lastly, we note that the polarity shifts we observe 
 differ from the local runs of \citet{Riols2018} which show continued positive toroidal fields around the disk midplane with vanishing fields above two disk scale heights.
 This discrepancy no doubt originates from the different setups and models. In particular, the vertical boundary condition is the likely culprit here; \citet{Riols2018} demonstrated the sensitivity of some elements of the dynamo to the boundary conditions. \citet{Riols2017,Riols2018} enforced $B_x=B_y=0$ and $dB_z/dz=0$ at $\pm 3H$ (the vertical boundary) which, by 
construction, cannot model field dragged by the velocity rolls  above/below $\pm3H$ \citep[see also][Fig. 9]{Shi2014}. In our Lagrangian simulations, which are global and do not require explicit boundary conditions, fluid elements can be followed well beyond
$\pm3H$. However, we acknowledge that the higher the altitude, the fewer are the computational elements, and thus the less accurate the numerical method. 

\begin{figure}[ht!]
\epsscale{1.}
\plotone{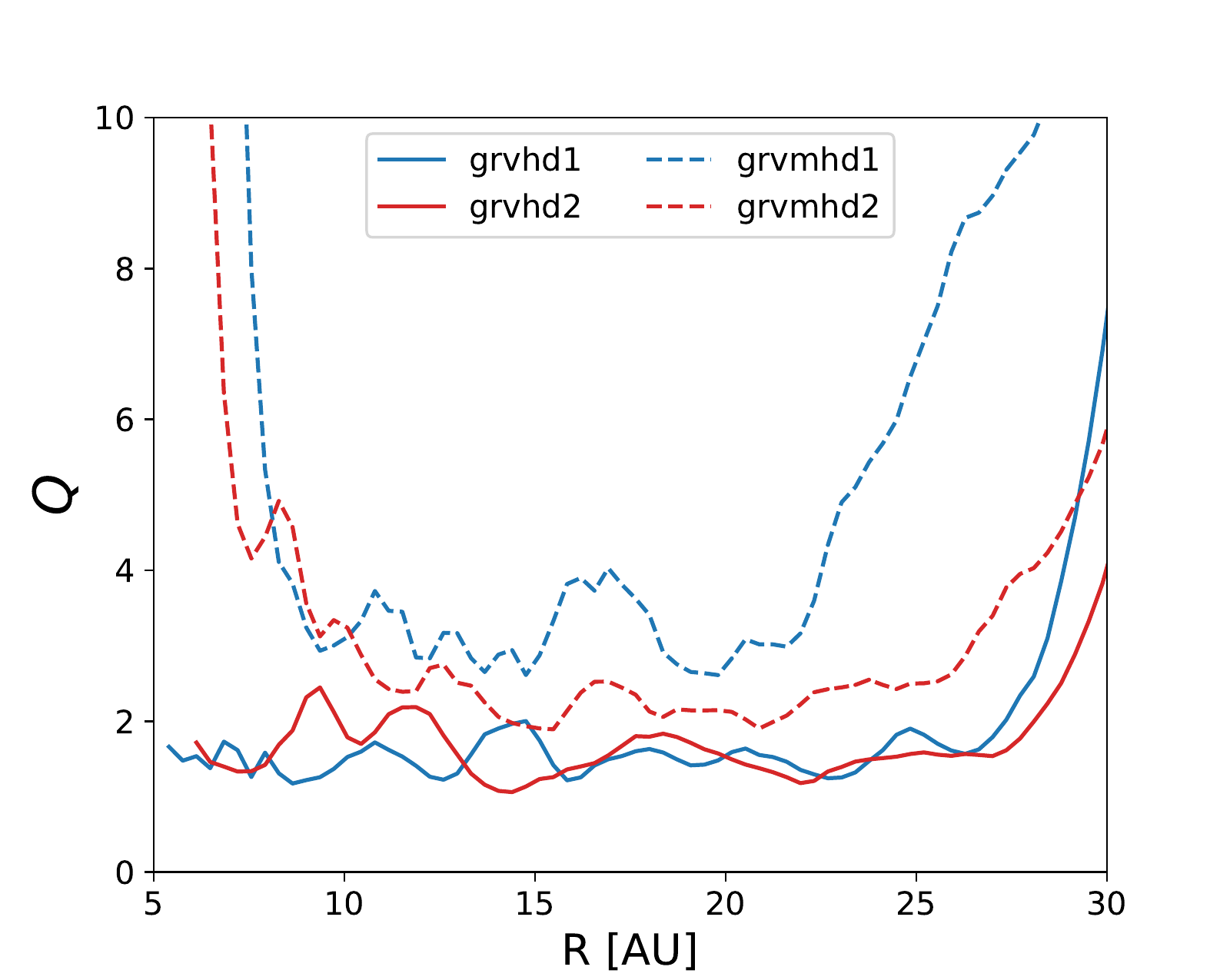}
\caption{The profile of the 2-dimensional \emph{Q} maps (see equation~\ref{eq:q2d}). The two MHD simulations saturate to higher \emph{Q} values than their HD counterparts. Note the azimuthally averaged \emph{Q} values are not weighted by density.\label{fig:qprofile}}
\end{figure}


\subsection{Magnetic field back-reaction on gravitational instability}
\label{sec:spirals}

\begin{figure*}[ht!]
\epsscale{0.9}
\plotone{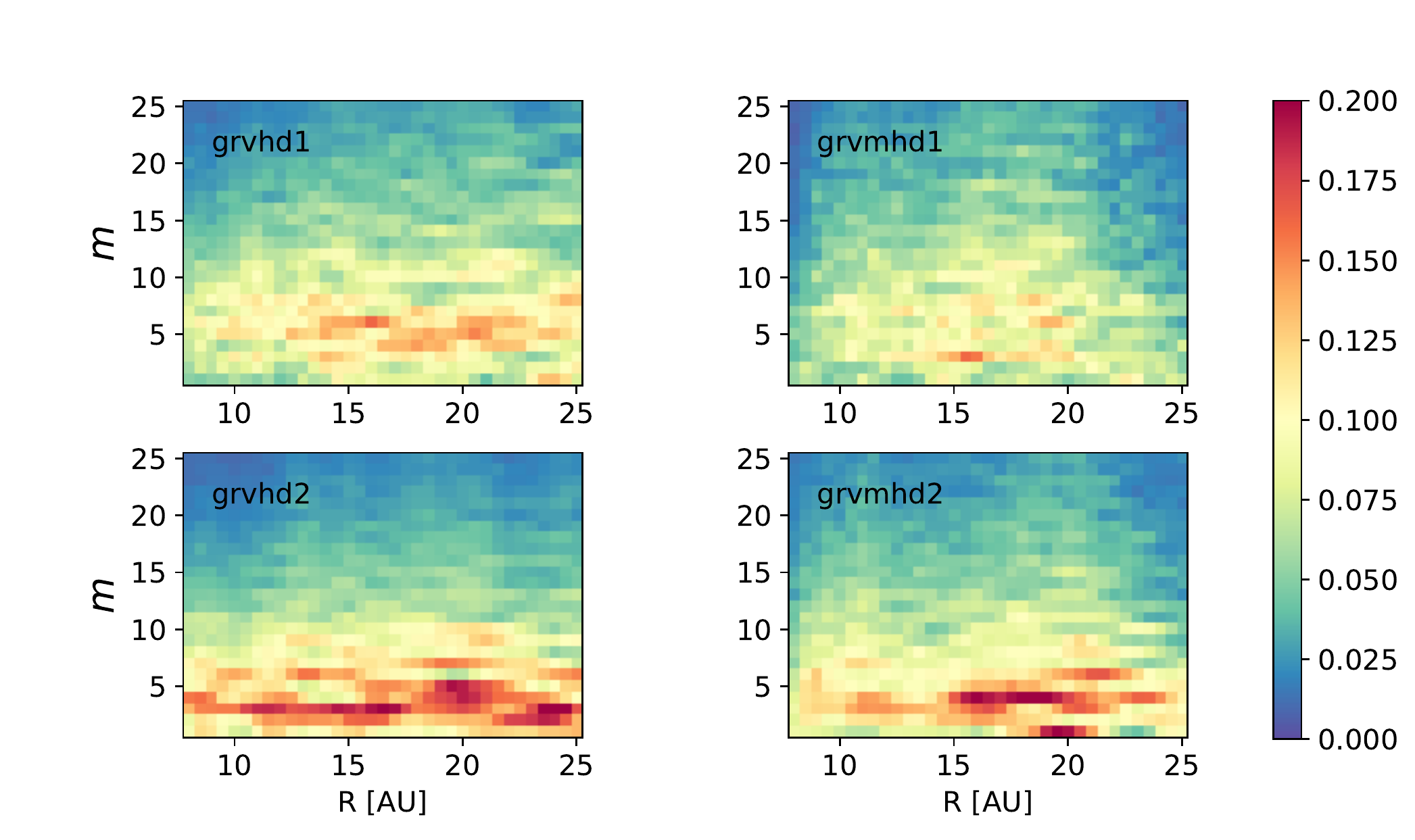}
\caption{Time averaged Fourier amplitudes of the density field in the different GI runs, with and without MHD. Radial cutoff is applied to avoid boundary condition effects\label{fig:fmap}}
\end{figure*}

The quasi-steady states described by our GI-MHD runs correspond to the situation when the dynamo has completed its `kinematic' phase and entered its saturated `nonlinear' phase, i.e. the magnetic field has grown to such a level that the Lorenz force is strong enough to react back on the flow that birthed it, and it will react back in such a way to halt the growth of magnetic field. In this subsection, we analyse the density field focusing on the effects of MHD turbulence on the spirals in this subsection. 

As well documented in the literature on the subject \citep{Durisen2007}, 
self-gravitating disks saturate to a state where the prescribed cooling is balanced by the heating due to the instabilities.  As shown in many previous studies \citep{Mayer2004,Rice2005,Cossins2009}, a GI disk saturates to a state with the Toomre \emph{Q} hovers around unity
($Q \sim 1-1.4$). We checked the 2-dimensional Toomre \emph{Q},
at the end of the SG simulations. The dense spirals have smaller \emph{Q} values than the dilute inter-spiral regions. The \emph{Q}  parameter can be as small as 0.4 in parts of the spirals with no fragments formation in both HD and MHD simulations.


We plot in  figure \ref{fig:qprofile} the profiles of the \emph{Q} maps in the various saturated states. The HD simulations have \emph{Q} hovering around 1.5 regardless of the disk mass (the disk star mass ratio is small here). However, grvmhd2 has $Q\sim 2.2$ and grvmhd1 has an even larger $Q\sim 3.2$. The extra heating from the dissipation of magnetic energy is likely responsible for the larger value of \emph{Q} at saturation, which was also observed in \citet{Riols2016, Riols2017}. The GI-MHD simulations are simply hotter and vertically more extended than their HD counterparts: the lower row of panels in figure \ref{fig:compile} makes this especially obvious. More specifically, \emph{Q} is larger in the inter-spiral regions of the MHD simulations than the HD simulations.

In the density maps of figure \ref{fig:compile}, we see that both GI and GI-MHD simulations exhibit strong spiral patterns with a rich mode structure. The GI-MHD simulations, though, show in general slightly more incoherent (`flocculent') spirals and, visually at least, smaller scale density fluctuations
are present compared to the corresponding HD simulations.



To gain a better quantitative insight in the density structure, we apply azimuthal Fourier transform to the volume density \citep[see, e.g.][]{Cossins2009}. The time-averaged mode amplitudes $\Sigma_m$ are shown at varying radius in figure~\ref{fig:fmap}. The hydro grvhd1 simulation is dominated by a $m=6$ mode throughout radii beyond $R=13$AU. \citet{Cossins2009} found similarly a dominant $m=5$ mode in disks about $0.1$ times the star mass with $\Sigma \propto R^{-3/2}$. When MHD is added any global radial coherence vanishes; at best there is the signature of an $m=3$ mode between 15 and 20 AU. It could also be argued that there is a redistribution of power to higher $m$.

On the other hand, the hydro run grvhd2 is dominated by a coherent $m=3$ mode throughout its entire radial extent. The dominant azimuthal mode \emph{shifts} however when MHD is added: in grvmhd2, the $m=4$ mode becomes the most prominent, as shown in figure~\ref{fig:fmap}, and there is less continuity in this dominance across different radii. There is hence a loss of global coherence. 

\begin{figure*}[ht!]
\epsscale{1.1}
\plotone{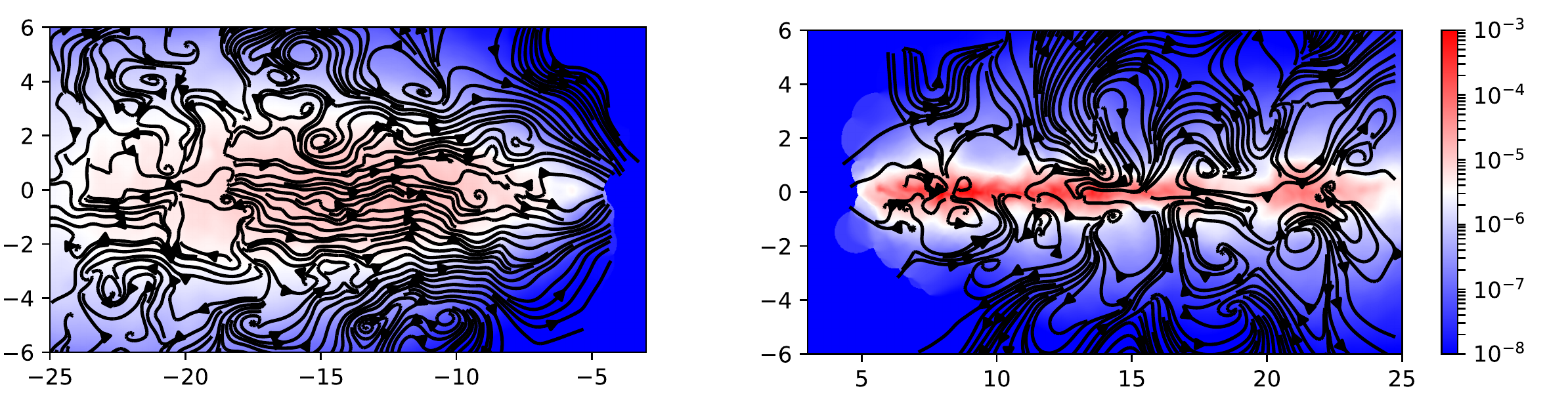}
\caption{Zoom in side-on density map of MRI-hr (left column) and grvmhd2 (right column) with over-plotted velocity streamlines; the full disks extend to $\pm$10 AU vertically (see figure~\ref{fig:compile}).    Many roll structures develop around the densest spiral center in grvmhd2 while no large scale motion in MRI-hr is observed. \label{fig:dynamo}}
\end{figure*}

\begin{figure*}[ht!]
\plottwo{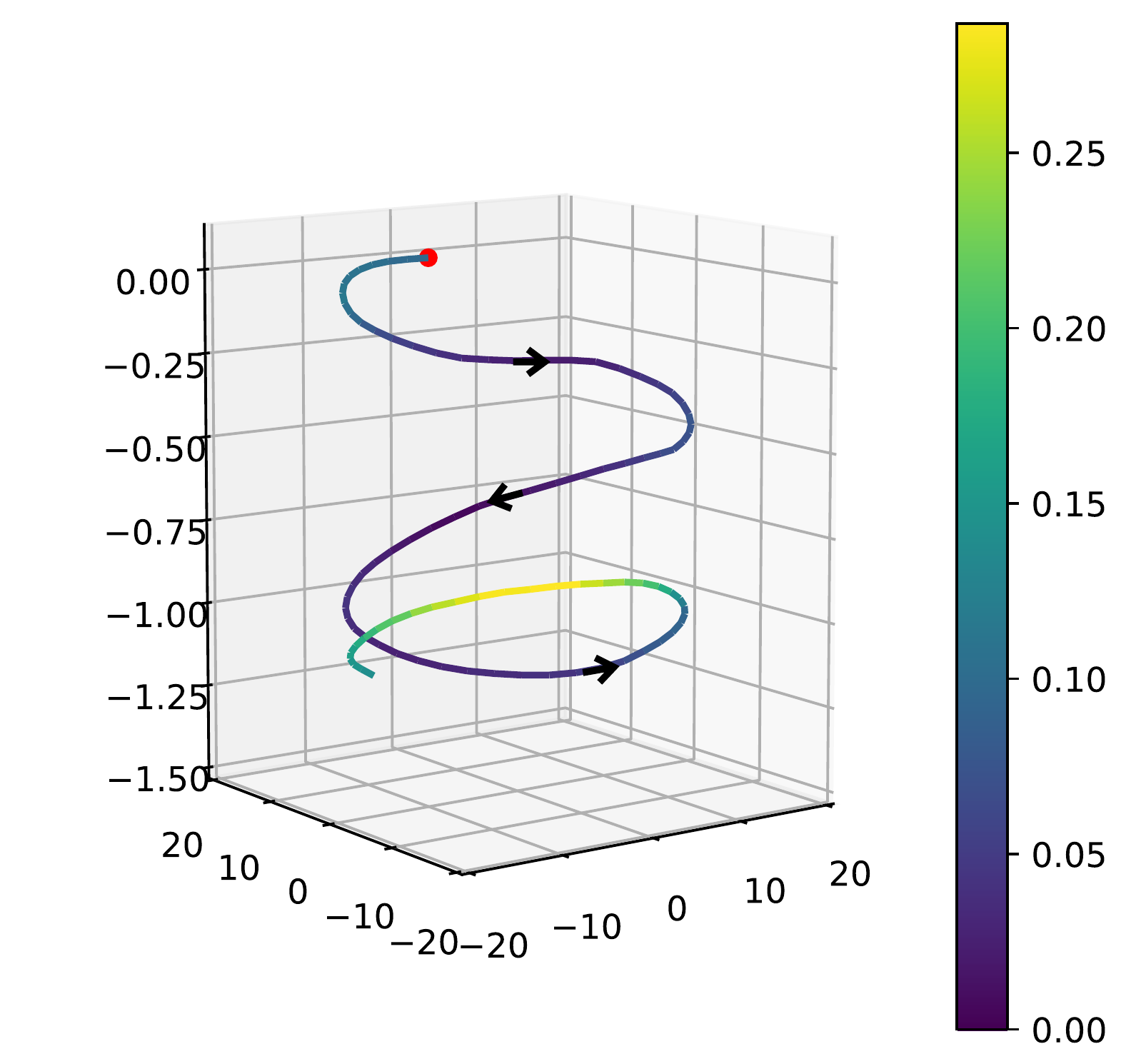}{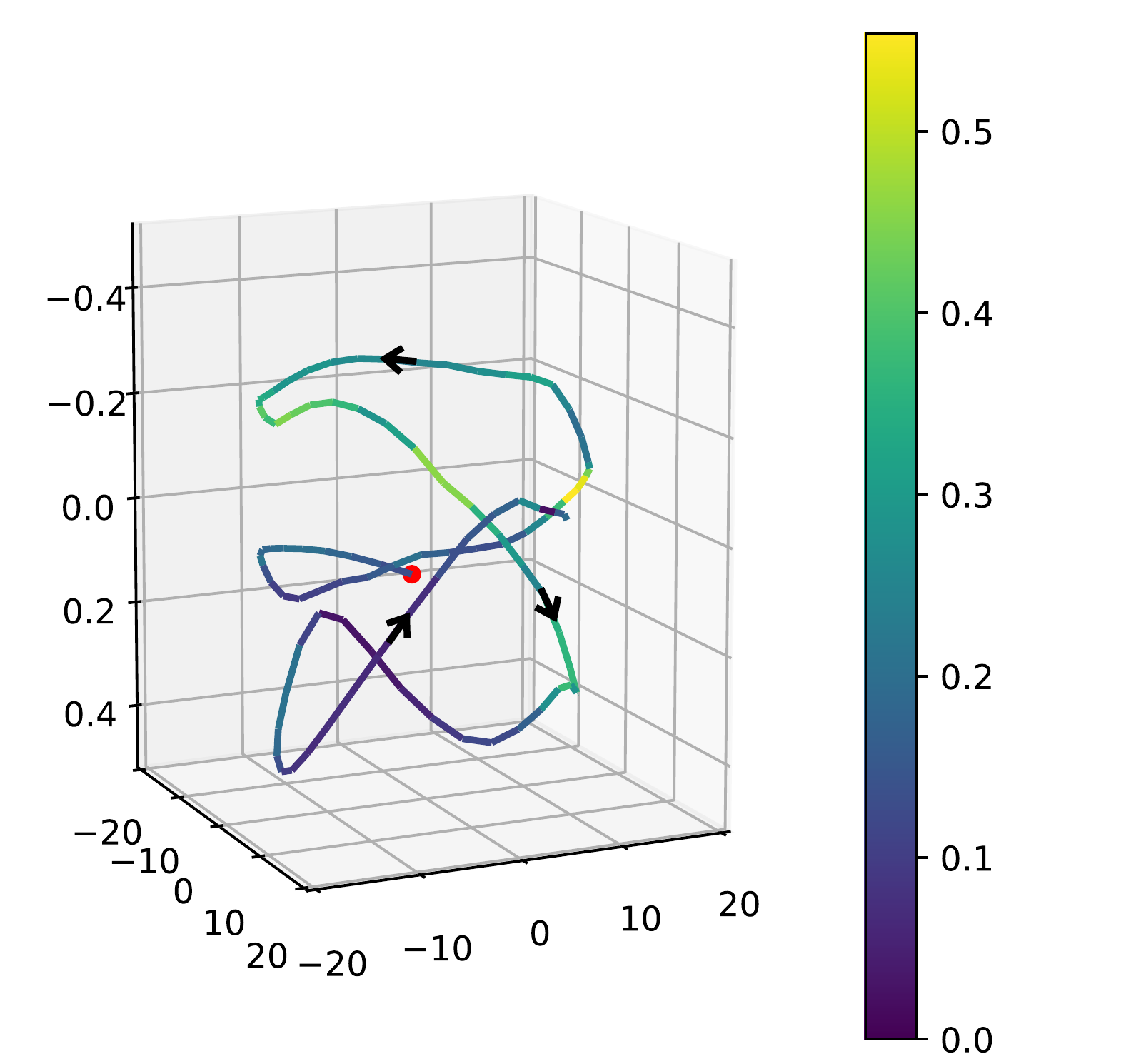}
\caption{ 3D trajectory of representative particles in the MRI-hr run (left) and in the grvmhd2 run (right), color coded in magnetic field strength. The star is at (0,0,0) and  the particles start from the red dots 16 AU away from the star followed for 2 local orbits. The magnetic field appears to be amplified when the
fluid element cross the midplane in the GI-MHD case, consistent with the notion whereby vertical rolls play a crucial role
in the magnetic dynamo generation. No such behaviour is present in the MRI case, indeed the randomly chosen particle does not even cross the
midplane. \label{fig:trace}}
\end{figure*}

It is noteworthy that the $m=4$ mode is particularly strong in grvmhd2 between 15-20 AU in figure~\ref{fig:fmap}. Following up on this, We measure the $m=4$ spiral pattern speed. Assuming a density perturbation $\propto e^{i\{m(\phi-\omega t)+kR+\phi_0\}}$, then the phase angle is $\phi_0-m\omega t+kR$, where $k$ is the radial wavenumber. The pattern speed between 15-20 AU of the major $m=4$ mode exactly equals the rotational angular speed at $16$ AU. This is also strictly true at different times. As a result, corotation resonance occurs at $\sim$16 AU as indicated by the stresses, which we show later in figure \ref{fig:stress}. The dominant modes in grvhd1/2 rotate significantly slower with a pattern speed close the orbital angular speed at 22 AU.


\subsection{Magnetic Dynamo in self-gravitating disks}
\label{sec:dynamo}

In this subsection we take a closer look at the process of magnetic field generation. \citet{Riols2017,Riols2018}, using local 
finite volume simulations in shearing boxes, showed that  the vertical circulations \citep[see, e.g.][]{Boley2006b,Mayer2007,Riols2018rolls} that naturally accompany spiral waves can, in alliance with differential rotation, make a dynamo loop. It is
fundamentally different to the MRI dynamo~\citep{Lesur2008,Gressel2010},
and has a vertically global character, working on scales larger than the disk scale height. Moreover, strong Ohmic dissipation that would completely quench the MRI in fact enhances the dynamo \citep{Riols2017,Riols2018}.

\begin{figure}[ht!]
\epsscale{1.2}
\plotone{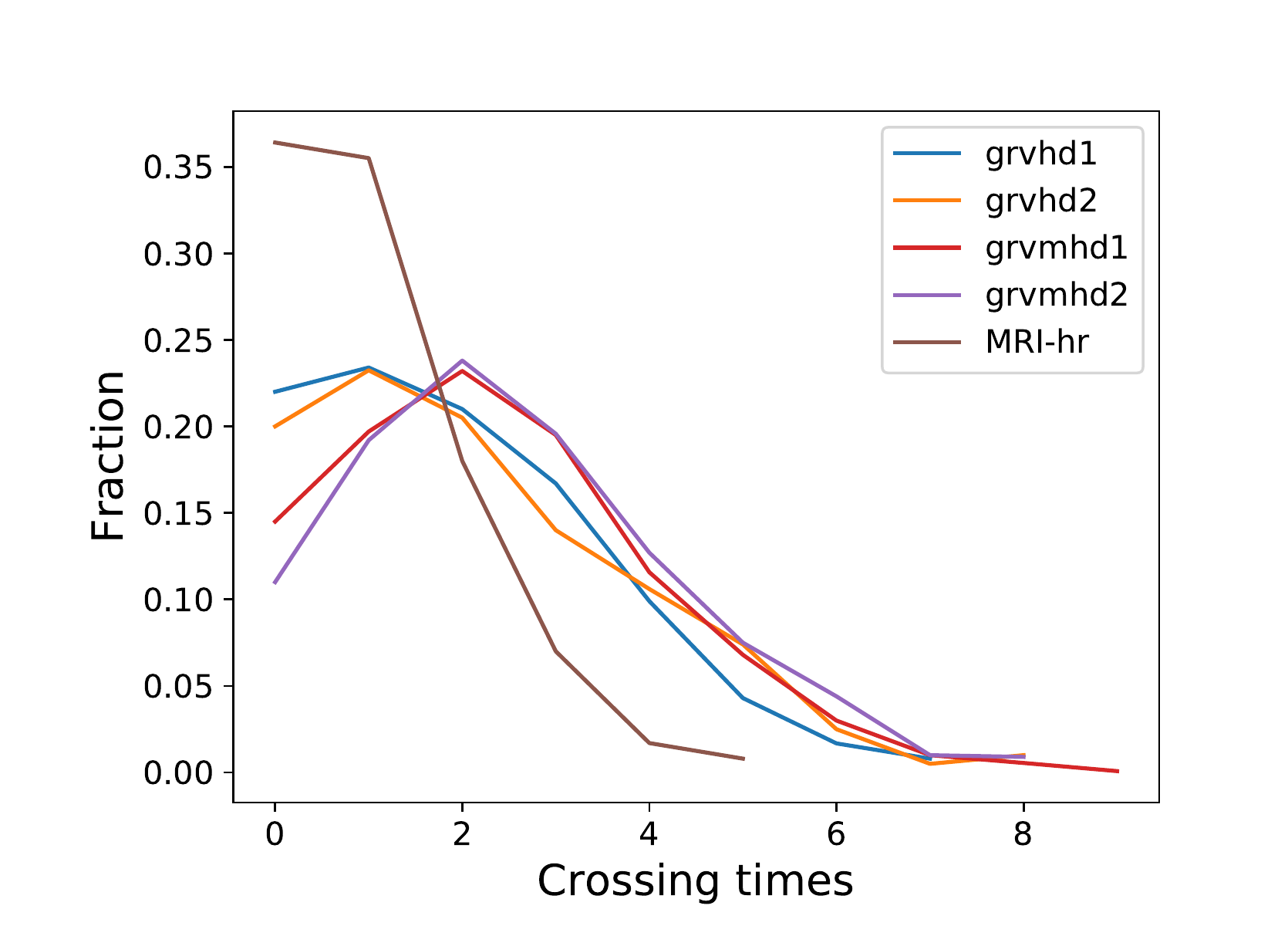}
\caption{ Distribution of midplane crossings for a subsample of representative particles in the saturated state for all runs. GI and GI-MHD
runs have similar distributions, with several midplane crossings per particle being typical, consistent with major vertical circulation
crossing the midplane, while this is rare in the MRI run, where most fluid elements never cross the midplane. \label{fig:cross}}
\end{figure}

In our global GI-MHD simulations we observe an analogous amplification of the magnetic field in conjunction with vertical circulation around spiral arms.
In figure \ref{fig:dynamo},  
 we show poloidal velocity streamlines and the volumetric density for the two simulations MRI-hr and grvmhd2. The latter clearly shows vertical velocity rolls, and these are correlated with the radial midplane structure. This is the crucial ingredient in the GI dynamo. In contrast, the MRI run does not exhibit such velocity structure. To make sure that these circulations did not issue from convection, we computed the squared vertical buoyancy frequency in grvmhd2 and found it always positive (as it was also in the corresponding HD runs).


In order to ascertain further that the latter interpretation of the magnetic field amplification really captures the physical behaviour
of the flow we exploited the Lagrangian nature of the code and followed individual trajectories of random fluid elements at 
arbitrary times. The typical trajectory is shown in figure \ref{fig:trace}, contrasting the MRI-hr and grvmhd2 runs. 
First, the different nature of the fluid elements' motion is evident, which reflects the different nature of turbulence in the two
regimes. Second, and most importantly, in the grvmhd2 runs the amplification of the magnetic field appears to occur as expected
as the fluid elements are dragged out of the disk midplane and then fall back.


To obtain statistically sound results we  count the midplane crossing times within 2 local orbits at 16 AU for $\sim$1000 particle that lie closest to the middle plane. Particles are chosen this way so that both the dense spiral and dilute inter-arm regions are sampled and increasing the number of particles shows converged results. We intentionally exclude the first immediate crossings that due to the small vertical motion of these sample particles (close to the midplane initially). In figure \ref{fig:cross}, we show the distribution of disk midplane crossings in the different runs, which highlights how in GI simulations, both with and without MHD,
fluid elements cross the midplane (from the top or bottom, as we do not distinguish from where in our statistics) much more frequently
than in the MRI case. Frequent midplane crossing reflect the action of the vertical rolls, which are focused around spiral arms, whereas in MRI 
vertical gas motion has no preferred accumulation point. We also note that vertical rolls in the local simulations of \citep{Riols2018} are not crossing the midplane, rather are confined within half the disk scale height.  Finally, from figure \ref{fig:cross} there is also marginal
evidence that rolls are more vigorous (higher crossing frequency) in MHD runs, probably reflecting some feedback loop effect of the
magnetic field onto the fluid circulation.




\subsection{Turbulent transport}


In the previous sections we have concentrated on characterising the growth of magnetic field and its influence on gravitoturbulence. We have employed relatively idealised numerical experiments as a platform to understand what is a fundamental physics problem. That being said, these simulations, even if idealised, may also bear on more concrete astrophysical applications, such as the issue of mass accretion and outflows in young stellar systems. Though missing many important physical effects (e.g. realistic radiative processes, ambipolar diffusion, the Hall effect, realistic ionisation profiles, etc),
the GI dynamo, as simulated here, could impact on how we understand accretion and outflows to work.

We first compare the efficiency of mass transport through the disk by turbulence generated in MRI, GI and GI-MHD runs, respectively. In Table \ref{t:stress}, to get a rough idea of magnitudes
we find that the magnetic stress, absent in purely hydrodynamic runs, dominates the gravitational and Reynolds stress in grvmhd1, and is comparable to the gravitational stress in grvmhd2. From these
number we see immediately that accretion should be greatly enhanced via the inclusion of magnetism: the GI dynamo creates strong correlated fields that transport angular momentum via the Maxwell stress. This is in agreement with previous local simulations. 
In Table \ref{t:stress}, the disk mass lost rate (accretion+outflow) in grvhd1/2 more than doubled after including MHD. This is perhaps one of the more exciting results of our simulations: magnetic fields enhance accretion in gravitoturbulent disks and thus speed up the evolution of young protostellar disks. However, we caution that whether the outflow (clipped particles) falls back onto the disk is uncertain. 

\begin{deluxetable}{c|c|c|c|c|c|c}[ht!]
  \tablenum{2}
  \tablecaption{Stresses and mass variation rates \label{t:stress}}
  \tablewidth{\textwidth}
  \tablehead{
    \colhead{Run } & \colhead{$H_{r\phi}$}& \colhead{$M_{r\phi}$} & \colhead{$G_{r\phi}$} & \colhead{$ \alpha$ }& \colhead{Accretion} & \colhead{Outflow}}
  \startdata
  MRI-hr      & 0.003     & 0.014& NA & 0.017       & 2.11 & 5.18 \\ 
  grvhd1       &  0.011    & NA  & 0.083  &  0.094  & 1.79     &  0  \\   
  grvhd2       &  0.012    & NA & 0.098&  0.110     &   6.61  & 0 \\
  grvmhd1    &  0.034      & 0.130& 0.063  &    0.227         &2.75 & 3.53 \\
  grvmhd2  & 0.035      & 0.080 & 0.092  &  0.207     &7.36 &5.89\\
  \enddata
  \caption{The stresses are the arithmetic average of the corresponding profile in figure \ref{fig:stress}. The accretion and outflow rates are in unit of 10$^{-6}M_\odot$/yr  measuring the mass variation (the last 1000 code units, figure \ref{fig:me}) due to accretion and clipping in the regions beyond 10AU. We note that the MRI disks are significantly hotter than GI-MHD disks because no cooling is applied (Table \ref{t:simulations}).  We caution that the low density disk surface is worse resolved than the midplane due to the adaptive nature of Lagrangian methods. }
\end{deluxetable}

\begin{figure*}[ht!]
\plotone{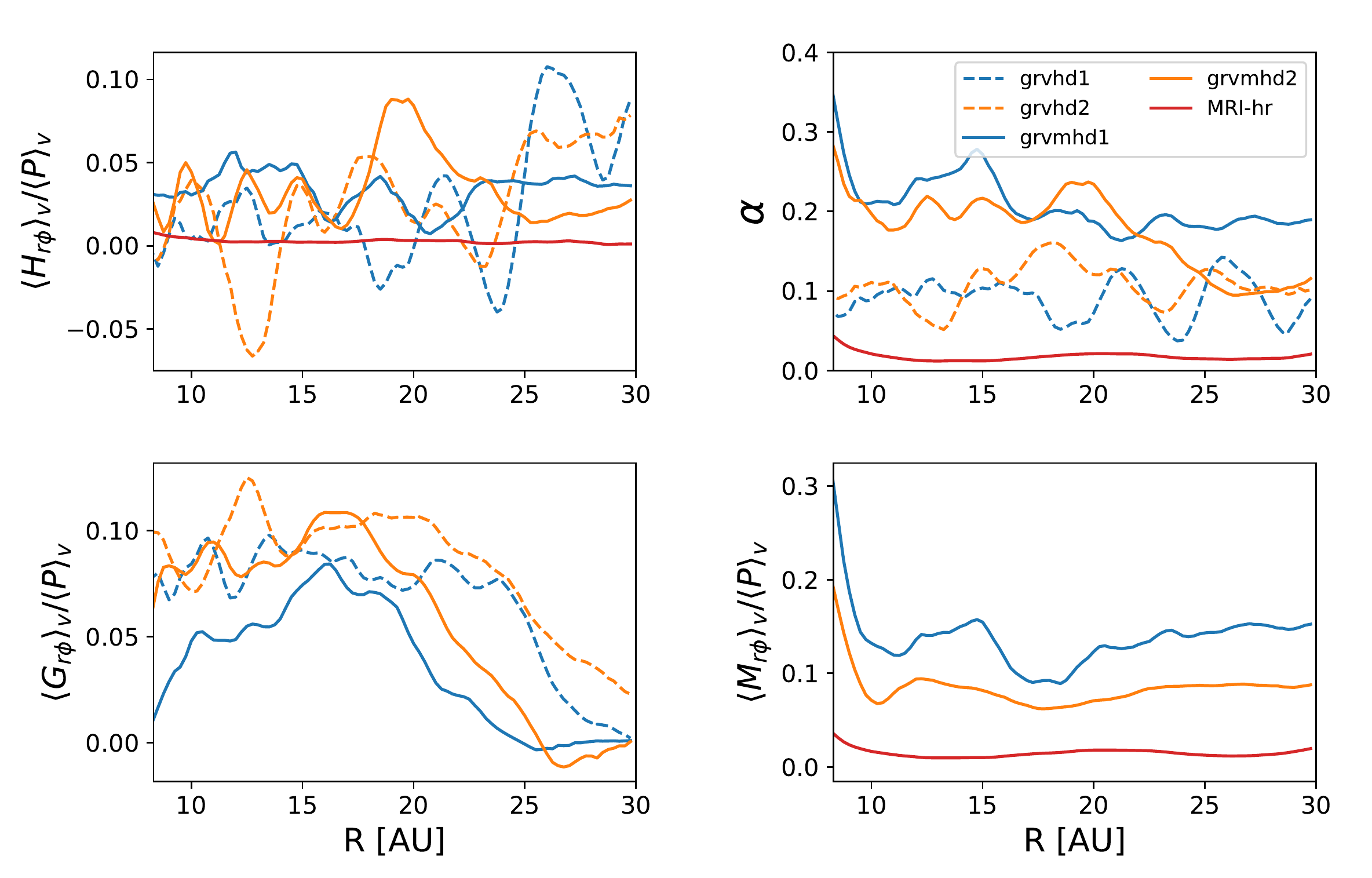}
\caption{Time averaged (thick lines in figure \ref{fig:me}) stresses. $\langle \alpha \rangle_V $=0.094, 0.110, 0.227, 0.207, 0.017 for grvhd1, grvhd2, grvmhd1, grvmhd2, MRI-hr respectively. The large Maxwell stress leads to the increase of  $\langle \alpha \rangle_V $ in the GI-MHD simulations compared to the GI simulations (Table \ref{t:stress}). \label{fig:stress}. }
\end{figure*}

We also computed the time-averaged  stresses (averaged over the last 1000 code time units) as a function of radius and plotted these in figure \ref{fig:stress}
for the different simulation runs.
 The Reynolds stress in GI fluctuates with radius and possesses negative values at some radii, and tracks the quasi-steady spiral structure. However, its contribution to the total stress is negligible in the simulations without magnetic field \citep[see also][]{Shi2014,Booth2018}. Interestingly, the Reynolds stress increases in the GI-MHD simulations, with an averaged value 0.034 for grvmhd1 and 0.035 for grvmhd2. The alteration to the basic gravitoturbulent flow is achieved via the Lorenz force and must be part of the dynamo saturation.
 
 On the other hand, the saturated dynamo blurs the spirals (as discussed in Section \ref{sec:spirals}) and decreases the gravitational stress in grvmhd1. However, the averaged gravitational stress in grvmhd2 equals that in the corresponding hydrodynamical simulations' because the spiral pattern remains strong even
in presence of the magnetic field. Note that the gravitational stress is small within the $R<10$AU and $R>20$AU regions due to low resolution and short simulation duration, respectively.
 

\begin{figure*}[ht!]
\epsscale{1.2}
\plotone{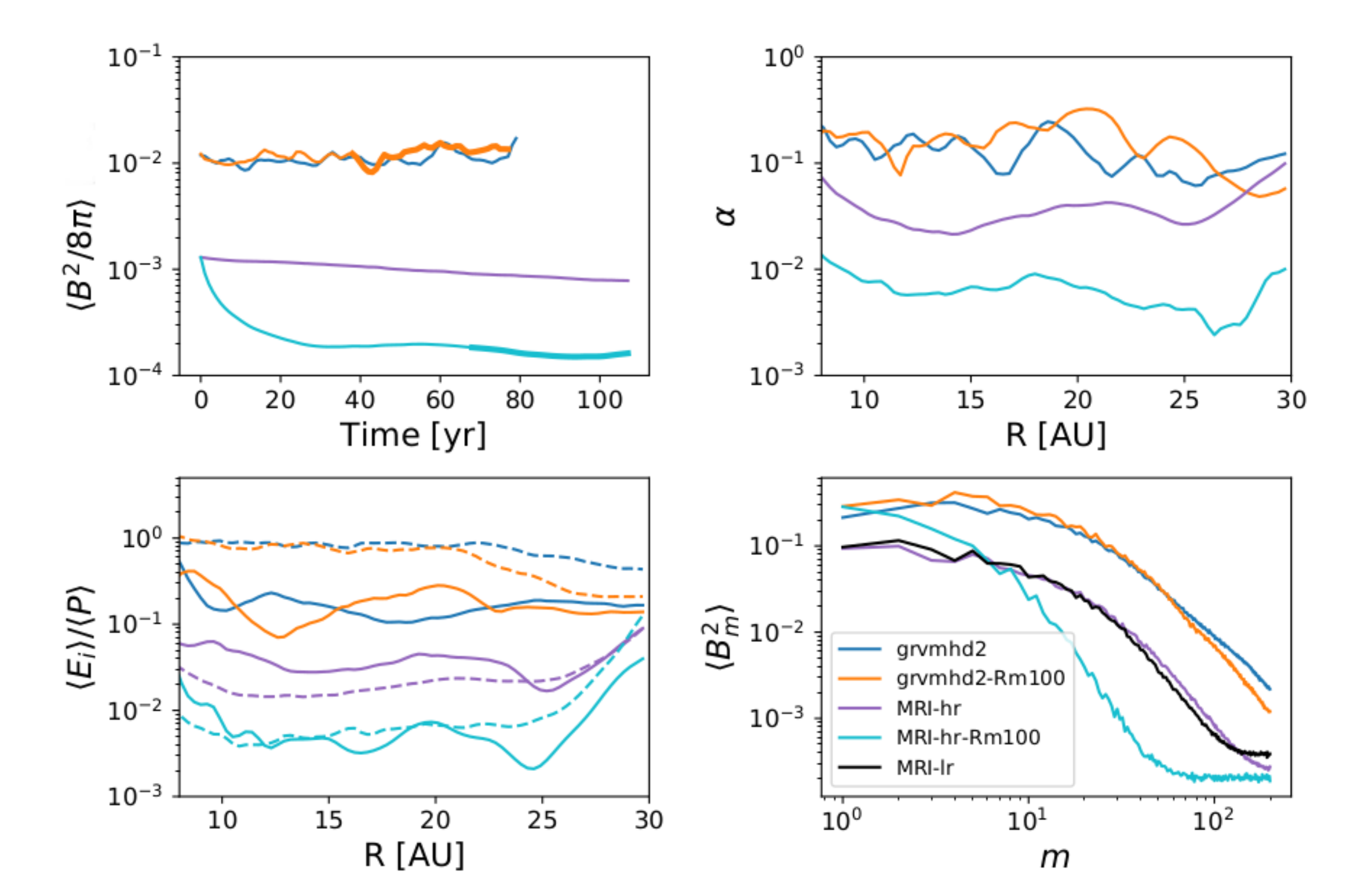}
\caption{Non-ideal MHD simulations of MRI-hr and grvmhd2 with Ohmic dissipation, $R_m$=100. The line series share the same legend as noted in the lower right panel; the dashed and solid lines in the lower left panel marks the kinetic and magnetic energy. Upper left: magnetic energy evolution, the thick line are the time span where the time average in the rest three panels are taken. Upper right: the $\alpha$ parameter. Lower left: the normalised average kinetic and magnetic energy. Lower right: the magnetic energy power spectra calculated similarly to figure~\ref{fig:bm}\label{fig:Rm100}}
\end{figure*}
 
  The radial averaged Maxwell stress (0.080) is comparable with the gravitational stress (0.092) in grvmhd2. Grvmhd1 has a higher averaged Maxwell stress of 0.13 with a smaller gravitational stress of 0.063. MRI-hr has a weaker Maxwell stress than both grvmhd1 and grvmhd2, which reflects the weaker amplification of the magnetic field relative to the runs in which the dynamo operates.
We also note that, $\alpha_M=2\langle M_{r\phi}\rangle /\langle B^2 \rangle \sim 0.4$ in MRI-hr, this being characteristic of resolved MRI simulations~\citep{Parkin2013,Hawley2011,Deng2019}.

We finish this section by pointing out that enhanced accretion witnessed must come at an energetic cost, as it were. Accretion liberates orbital energy and transforms it into heat; thus faster accretion leads to faster heating, and yet the cooling timescale is \emph{held fixed} between hydro and MHD runs. In order to achieve a steady state there must be another source of cooling, and that here is achieved through an outflow. As Table \ref{t:stress} shows both grvmhd1 and grvmhd2 exhibit significant loss of mass vertically. This wind in itself is worthy of close study, not least because it might connect to observed wide-angle low
speed molecular outflows from evolved class 0
and I objects \citep{Bally2016}. For now we merely point it out and also caution that its dependence on the numerical particulars of the simulations requires further exploration.

\subsection{Non-ideal MHD effects: Ohmic resistivity}
\label{sec:ohmic}

Non-ideal MHD effects can often be important in astrophysical disks, and is thought to suppress the MRI for many if not most radii in protostellar disks ~\citep{Blaes1994,Sano2000,Balbus2001,Kunz2004, Bai2013,Lesur2014,Bai2014,Gressel2015}. They also play an important role in the early stages of disk formation as their role
in crucial in avoiding the angular momentum catastrophe caused by magnetic breaking and allow an extended disk to form \citep[e.g.,][]{Li2011,Wurster2018}.  Ambipolar diffusion plays an important role in this context, and is generally considered the
dominant effect in the outermost regions of disks \citep[e.g.,][]{Mellon2009}, while the Hall effect has been found to have a
potentially important effect on the size of the disk that results from molecular cloud collapse depending on the relative orientation 
of the magnetic field and the spin axis \citep{Krasnopolsky2011,Marchand2018}.
However, at very high densities, such as in the midplane of the disk, in the gravitationally unstable region (R $> 10$ AU) where
MHD turbulence is seen to grow via the spiral dynamo mechanism, Ohmic resistivity may play a strong and possibly dominant part in the dynamics.

We present preliminary results with magnetic resistivity which confirm the robustness of the spiral-wave dynamo 
mechanism, as shown earlier in local boxes \citep{Riols2018}.  There the dynamo was shown to occur even in strongly Ohmic environments ($R_m=c_s^2/\eta\Omega\sim 1)$. To test the role of Ohmic dissipation we add in explicit magnetic resistivity in grvmhd2 at 320 yrs.  We rerun the MRI-hr simulation with $R_m=100$ throughout the disk as a comparison study.

With such a strong magnetic diffusivity, the magnetic energy of the MRI turbulence quickly decays (figure~\ref{fig:Rm100}).
In contrast, the spiral wave dynamo continues, after an initial readjustment period. We calculated time averaged values for one orbit period at 16 AU (thick lines in figure \ref{fig:Rm100}) for the simulations with Ohmic dissipation. Although the averaging time span is smaller than the previous ideal MHD cases, the disks are saturated when we start the time averaging. The effective viscous stress in MRI drops more than an order of magnitude. However, the energy partition and stress in grvmhd2 is only slightly affected.

We also plot power spectra of the magnetic energy. Firstly, we see that addition of Ohmic resistivity completely alters the spectrum of the pure MRI run, as might be expected. On the other hand, the spectrum of the Ohmic and ideal grvmhd2 runs are roughly similar. The inclusion of Ohmic resistivity does push power to longer scales (smaller $m$), in accord with \citep{Riols2018}, and the small $m$ tail becomes steeper, as the dissipative scale is longer.

\section{Discussion}
\label{sec:disc}

Our global  GI-MHD simulations, which are the focus of this paper, 
are still quite idealized as they adopt a simplified cooling prescription, assume the disk has some unspecified pre-existing ionization, and neglect to explore the role of two of the three
no-ideal MHD effects (the Hall effect and ambipolar diffusion).
This minimal setup is chosen to illustrate the basic physical
processes operating, and the differences with the better studied
and understood regimes of GI and MRI in disks.
We plan to build upon the current setups incrementally to account for more complex thermodynamics and non ideal MHD effects \citep{Wardle2007,Lesur2014}.

The strength of the spiral density waves, and  thus its associated dynamo, is sensitive to the cooling rate in self-gravitating disks. We expect the spiral-wave dynamo dominates MRI, in the ideal MHD regime, at least when $\tau<20\Omega^{-1}$ \citep{Riols2017}. The cooling rate of early stage disks is uncertain. In pure HD simulations with radiative transfer, \citet{Boley2006a} found $\tau \sim 20 \Omega$ around 20 AU.  In our MHD simulations, the internal energy/temperature in the disk corona is about 20 times larger than the disk mid-plane temperature ($\sim 10$ K)  while such hot disk corona is absent in the HD simulations. The hot corona and associated vertical circulation may help to cool the disk efficiently \citep{Boley2006b}. Further studies with radiative transfer~\citep[see, e.g.][]{Mayer2007} is necessary to address the cooling rate of the GI-MHD turbulent disk. We also note that infall can do the same job to bring the disk into GI similarly to the \emph{ad hoc} cooling~\citep{Boley2009}.

The ionization and chemical state of the disk is crucial for any disk process
involving magnetic fields, and the mechanism proposed here is no exception.
The disk temperature in our MHD simulations, which ranges from 10 to $100-200$ K, is insufficiently  high to provide any appreciable thermal ionisation, though it might be interesting to ask what temperatures can be reached in spiral shocks; a four-fold increase in temperature will bring the collisional ionisation of the alkali metals into play, and a steep rise in the ionisation fraction.   Past work on GI with 3D global simulations including radiative transfer and the complex roles of grain chemistry (e.g. porosity and ice coating) have shown that in spiral shocks the gas could heat significantly more than what is observed in this paper - up to several hundred K \citep{Podolak2011}. 

In any case, the ionisation source of most importance here must be non-thermal, arising from cosmic rays, irradiation by nearby OB stars, or from the central star itself. Each of these processes is somewhat poorly constrained, and for young disks the associated ionisation profiles have not been studied to the same level of detail when compared to their older T-Tauri relatives. It would be useful if future work could be dedicated to estimating these profiles.

The effect of the magnetic field and the dynamo-induced circulation on the critical regime of disk fragmentation will be studied in
a future paper. Expected effects are extra support against contraction into self-gravitating clumps from magnetic pressure support, and, conversely, dissipation of angular momentum of a gas clump in the disk close to contract and fragment, which would promote contraction. 
Which effect will dominate over the other and how the magnetic fields affect the mass of fragments can only be determined with high resolution simulations. Early attempts by \citet{Fromang2005} found fragments formation when small scale MHD turbulence starts to play a role although the fragments are dispersed later due to a lack of resolution. Recent SPH MHD simulation find that magnetic tension force increase the fragments mass in fast cooling regime while magnetic fields suppress fragmentation in slow cooling regime~\citep{Forgan2016}. However, they used a very low resolution (less than 3\% particles used here) which is definitely not able to resolve the MHD turbulence (see Appendix \ref{sec:res-test}). The suppression of fragmentation may also be caused by unphysical fields growth in stratified MHD disk simulations using traditional SPH MHD \citep{Deng2019}. The problem is still open and we will explore it next.

\section{Conclusions}
\label{sec:con}
We carried out three-dimensional global MHD simulations of self-gravitating accretion disks using the MFM method \citep{Hopkins2015a,Hopkins2015b} to study the interaction between
gravito-turbulence and a magnetic field (zero-net-flux) threading the disk.
For comparison, we also ran pure GI and MRI simulations using the same disk models and a similar numerical setup. The global MRI runs are meant to provide a point of reference and comparison for the GI dynamo runs to prove that the latter are a different phenomenon to the MRI. In fact, the difficulty particle codes have in describing the MRI \citep{Deng2019} is to our advantage here, as the MRI will be possibly weaker and less prevalent, letting us attribute magnetic growth to GI dynamo more confidently.  Our main findings can be summarized as follows:


1. We confirm that global GI turbulence efficiently generates strong magnetic fields, and thus acts as a dynamo. The field growth and saturation is quantitatively and qualitatively different to the MRI. First shown to occur in shearing boxes, this is the first demonstration that the dynamo also works in a global disk model. 

2. By examining the vertical circulations associated with GI spiral waves, we showed that some aspects of the dynamo mechanism proposed by \citep{Riols2017,Riols2018} appear in global disk simulations. The saturated field strength and toroidal field polarity variation is affected by the disk mass likely through different spiral patterns.  

3. We observe how the saturation of the dynamo impacts on the flow field: the disk becomes hotter via magnetic dissipation and, consequently, the Toomre $Q$ is  significantly larger than in the hydro gravitoturbulence; the GI spirals become less coherent and flocculent; and there develops a greater Reynolds stresses, and potentially lower gravitational stresses.

4. The dynamo enhances the total transport of angular momentum through the disc substantially, when compared to purely hydrodynamical gravitoturbulence. Thus magnetic fields can `speed up' the evolution of young protostellar disks.  Accompanying this accretion is an outflow, which expels energy from the disk; this phenomena needs to be studied further and its numerical particulars (in particular, robustness) understood.

5. Ohmic resistivity, while killing off the MRI, has little impact on the dynamo, except to transfer power to longer scales than otherwise. This is in accord with previous local simulations \citep{Riols2018}.

As emphasised, our simulations are quite idealised, in order to more clearly exhibit and identify the various physical processes underlying the dynamics. Future work, however, must begin to add, piece by piece, the most relevant physics for protostellar disks. We identify as the most urgent: (a) more realistic ionisation profiles and non-ideal MHD effects, and (b) different magnetic flux configurations, especially the case when the disk is threaded by a large-scale poloidal field. Both will open up the exploration of how gravitoturbulence, the GI dynamo, and magnetically mediated outflows interact.



\bibliographystyle{aasjournal}
\bibliography{references}


\appendix

\section{Resolution study}
\label{sec:res-test}
\begin{figure*}[ht!]
\plotone{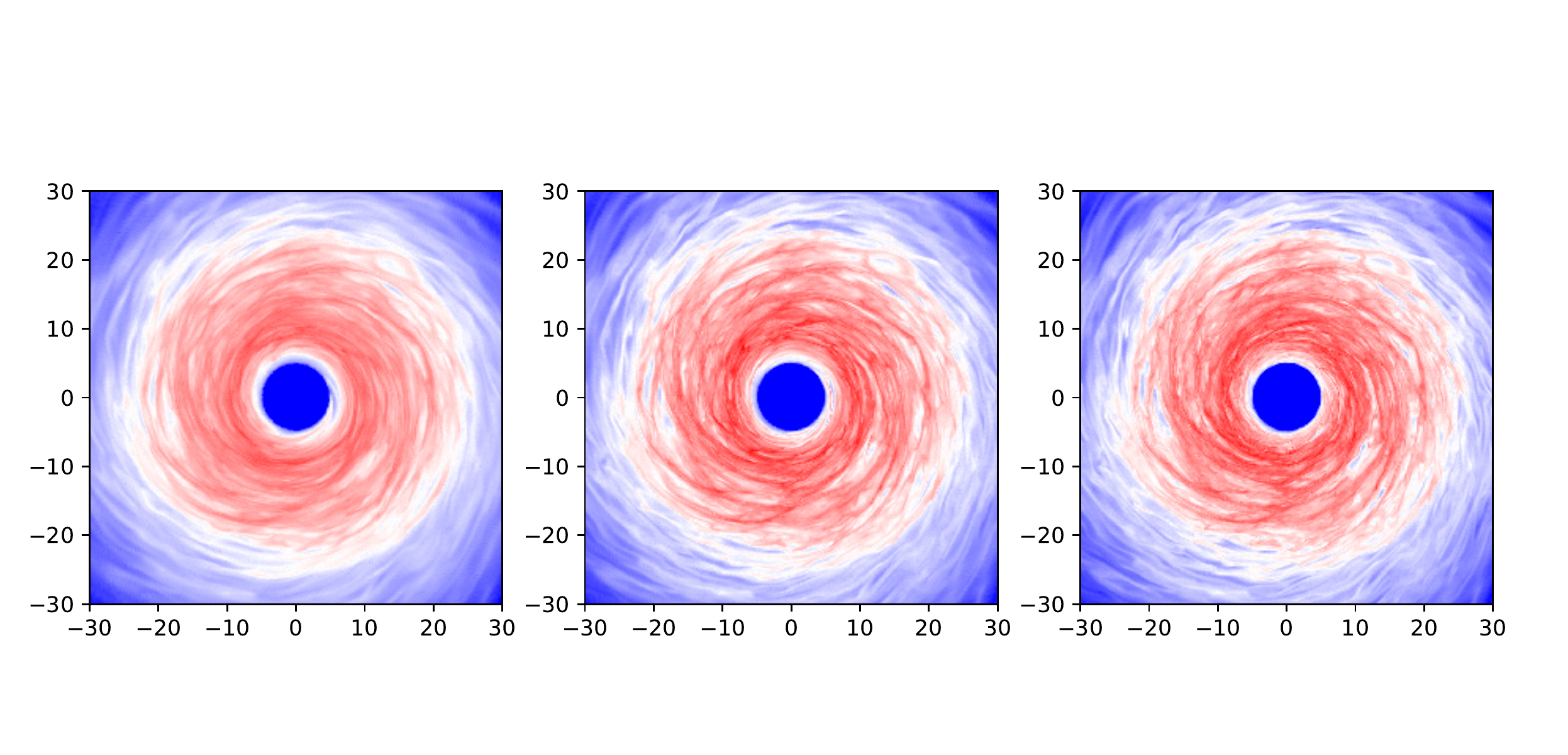}
\caption{Midplane density plots with the same color scheme in figure \ref{fig:compile}  at 820 yrs for the 18M, 36M and 72M particle simulations from left to right. The spirals remain almost identical after doubling the number of particles in our fiducial model while the lower resolution simulation smears out the flocculent spirals to some extent. \label{fig:spirals-res}}
\end{figure*}

To test the numerical convergence directly we reran grvmhd1 from 760 yrs and varying the resolution by merging or splitting particles. The tests were run for 70 yrs. We show the midplane density in simulations at the three different resolutions in figure \ref{fig:spirals-res}. The 18M particles simulation clearly blurs the spirals. 

\begin{figure}[ht!]
\epsscale{0.5}
\plotone{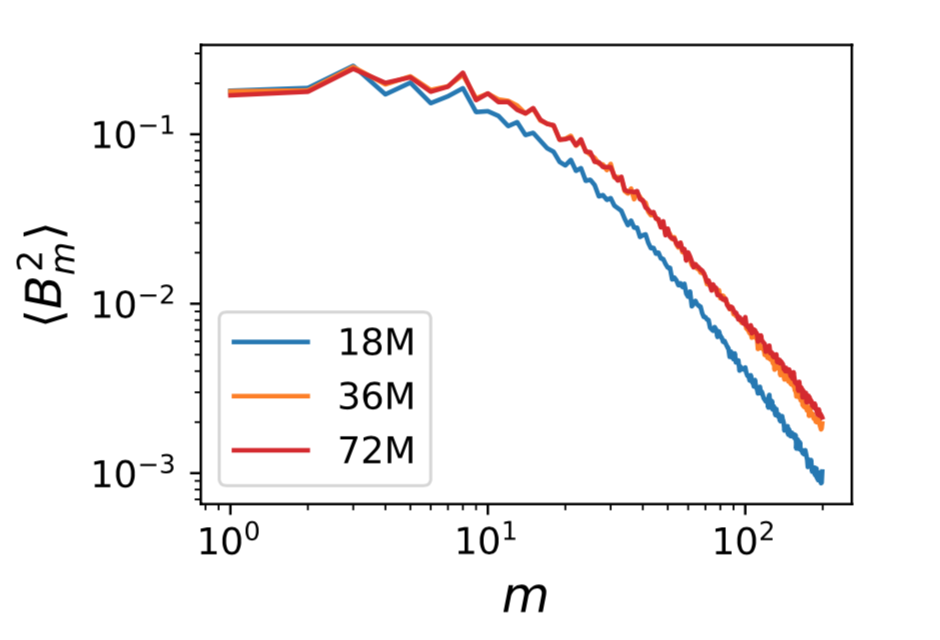}
\caption{The magnetic energy power spectrum averaged between 10-20 AU for the 70 yrs convergence test (similar to figure \ref{fig:bm}). The prediction of our fiducial model varies little after doubling the number of particles while halving the number of particles leads to much weaker turbulence. \label{fig:me-res}}
\end{figure}

The 18M particle simulation has weaker fields and smaller stresses in all three components and thus a smaller $\alpha$ than the two other simulations. The 36M particle simulation (the resolution adopted in our main paper) agrees well with the 72M particle simulation in stresses and magnetic energy. Especially, the space and time averaged magnetic energy power spectrum appears converged at 36M particles in figure \ref{fig:me-res}.

We note that it is hard to compare resolution in Eulerian and Lagrangian simulations due to the intrinsic adaptivity in the latter method. In global pure GI simualtions, at least when the disk is approaching fragmentation, around 10 times more computational elements are needed in grid codes than a Lagrangian method \citep{Durisen2007,Mayer2008}. As for the MHD module, we also do not have a uniform resolution scale in Lagrangian methods. We have a resolution scale \citep[not the smoothing length, see][]{Dehnen2012,Deng2019} 1/16 and 1/20 of the disk scale height in the spiral wave center near 10 AU and 16 AU in grvmhd1. The corresponding values for grvmhd2 are 1/22 and 1/28. 

We calculated the averaged quality factor, i.e., the number of effective cells per characteristic MRI wavelength \citep{Hawley2011,Deng2019} in our GI-MHD simulations to give a reference for future studies. In the saturated state of grvmhd1/2 , $\langle Q_\phi \rangle$ is about 30, 28 and $\langle Q_z \rangle $ is about 5, 5 respectively. We caution that the GI dynamo is on large scale and fundamentally different from the MRI dynamo so that these quality factor does not reflect how well the dynamo is resolved. 

\end{document}